\documentclass[aps,article,twocolumn,amsmath,amssymb,eqsecnum, showpacs]{revtex4}
\usepackage[colorlinks=true,citecolor=blue,linkcolor=blue]{hyperref}
\usepackage{graphics}
\usepackage{graphicx}
\usepackage{epsfig}
\usepackage{setspace}
\usepackage{amssymb,amsmath}
\usepackage{color}
\usepackage{balance}
\def \be {\begin{equation}}
\def \ee {\end{equation}}
\def \ba {\begin{eqnarray}}
\def \ea {\end{eqnarray}}
\def \bm {\begin{displaymath}}
\def \em {\end{displaymath}}
\def \br {{\bf r}}

\begin{document}
\title{ Structures of simple liquids in contact with nanosculptured surfaces}
\author{Swarn Lata Singh, Lothar Schimmele, and S. Dietrich}
\affiliation{\hspace*{5.8cm}{Max-Planck Institut f\"{u}r Intelligente Systeme,} 
\newline
{\centerline{ Heisenbergstr. 3, D-$70569$ Stuttgart, Germany and}}
\newline
$IV$. Institut f\"{u}r Theoretische Physik, Universit\"{a}t Stuttgart, 
Pfaffenwaldring, $57$, D$-70569$ Stuttgart, Germany}
\date{\today}
\begin{abstract}

We present a density functional study of Lennard-Jones liquids in contact with a 
nano-corrugated wall. The corresponding substrate potential is taken to exhibit a 
repulsive hard core and a van der Waals attraction. The corrugation is modeled 
by a periodic array of square nano-pits. We have used the modified Rosenfeld density 
functional in order to study the interfacial structure of these liquids which with 
respect to their thermodynamic bulk state are considered to be deep inside their liquid phase. 
We find that already considerably below the packing fraction of 
bulk freezing of these liquids, inside the nanopits a three-dimensional-like 
density localization sets in. If the sizes of the pits are  commensurate with the 
packing requirements, we observe high density spots separated from each other in all 
spatial directions by liquid of comparatively very low density. The number, shape, size, 
and density of these high density spots depend sensitively on the depth and width of 
the pits. Outside the pits, only layering is observed; above the pit openings these layers are 
distorted with the distortion reaching up to a few molecular diameters.  
We discuss quantitatively how this density localization is affected by the geometrical features 
of the pits and how it evolves upon increasing the bulk packing fraction. 
Our results are transferable to colloidal systems and pit dimensions corresponding 
to several diameters of the colloidal particles. For such systems the predicted unfolding
of these structural changes can be studied  experimentally on much larger length scales 
and more directly (e.g., optically) than for molecular fluids which typically call for 
sophisticated X-ray scattering.
\end{abstract}

\maketitle
\section{\bf Introduction}
The properties of liquids near solid surfaces are of interest for a wide range of 
technological and biological issues.  Wetting or non-wetting properties of solids, 
notably of structured surfaces, play an important role in applications -- such as 
coating, design of self-cleaning materials and of micro-fluidic devices, as well 
as tribology \cite{Wang, Blossey, Delamarche,Bhushan, David,Heuberger} -- and also in biology 
\cite{Neinhuis, Sharma}. In order to modify the wetting 
properties of solid surfaces, various methods can be used in order to tailor their
geometrical and chemical topography down to the nanoscale \cite{Yanagishita, Martines}. 
It turns out that even changing this topography only on the {\it nanoscale} leads to strong 
{\it macroscopic} effects such as a significant variation of the contact angle of 
sessile drops. Accordingly, in order to understand this peculiar amplification mechanism
one has to reveal the structure of liquids near patterned walls.

The geometric surface topographies studied most frequently in experiments are arrays 
of pillars or lamellae and arrays of pits or grooves carved into a flat and 
chemically homogeneous surface. Between pillars and within pits or grooves 
the fluid is strongly confined by solid walls. It is well known that the confinement 
of fluids as in slits or in cylindrical pores gives rise to effects like capillary 
condensation or capillary evaporation which can be understood based on thermodynamics, 
involving surface and interfacial tensions and bulk pressure. In nano-pores the 
presence of wetting films may lead to strong deviations from macroscopic predictions.
In order to incorporate these effects into a theoretical description one has to
resort to at least a mesoscopic treatment, e.g., based on an effective interface 
Hamiltonian  \cite{Dietrich}. Furthermore, in nanopores packing effects may become noticeable 
throughout the system, leading to pronounced layering. In order to include these latter 
effects into a theoretical description one has to use actual microscopic theories like molecular
dynamics, Monte-Carlo simulations, or classical density functional theories (DFT) with
sufficiently sophisticated functionals. A review of fluids in nanopores is provided in Ref. 
\cite{Evans}; for theoretical studies of capillary evaporation, capillary condensation, 
and wedge filling based on classical DFT with a focus on universal behavior see Refs. 
\cite{Roth1, Roth2, Malijevsky, Yatsyshin}. 
Concerning investigations using Monte Carlo simulations see Refs. \cite{Schoen1, Schoen2} 
and the review in Ref. \cite{Schoen3}.

Here we do not discuss fluids in complete confinement like provided by macroscopically extended 
slits or cylindrical capillaries as studied in the aforementioned literature. 
Instead, by using classical DFT we investigate one of the typical surface topographies 
used to manipulate wetting properties. We study a wall endowed with an array of pits which is in 
contact with a fluid of prescribed bulk number density. The fluid is homogeneous sufficiently far 
away from the wall and thermodynamically deep in the liquid part of the phase diagram, {\it i.e.},
far away from both liquid-vapor and  liquid-solid coexistence. We focus our investigations on narrow 
and shallow pits with depths and widths in the nano-meter range.
We expect strong confinement effects which, however, are expected to differ from
those in narrow but long capillaries. Our emphasis is on the structure of simple liquids 
in that given environment.  We do not study the Cassie-Baxter to Wenzel transition or capillary 
filling problems which have been addressed experimentally for wider pits and described in terms 
of macroscopic thermodynamics \cite{ Murakami, Martines, David} or by using mesoscopic 
approaches \cite{Miko1, Miko2, Hofmann}. There is also 
a DFT study devoted to wetting and filling transitions at substrates endowed with macroscopically 
long rectangular grooves \cite{Malijevsky2}. The fluids studied there are taken to be close to bulk liquid-vapor 
coexistence; accordingly the liquid structure per se is not analyzed.
Furthermore molecular dynamics studies have been performed concerning wetting of surfaces textured with 
nano-sized pillars \cite{Zhang, Jiang}. Also in this case the emphasis is on mesoscale structures, without 
discussing the liquid structure on the microscale.

Here we fill this gap by studying specifically the structures of confined liquids on the microscale. 
We aim at understanding whether on that scale strongly inhomogeneous liquid density distributions emerge 
and if so what their characteristics are. We analyze the dependence of these liquid structures on the pit 
dimensions and on the bulk packing fraction. It is particularly interesting to follow the changes of the 
liquid structure upon choosing the pit dimensions to be commensurate or incommensurate with the length 
characterizing packing effects of the local density distribution.
We are also interested in the liquid structures above the pit, with a view on how deep into the 
bulklike liquid traces of the specific structures forming within the pits are still detectable
on the outside.

Our studies are somewhat related to capillary freezing; for corresponding experimental 
and theoretical studies see Refs. \cite{Heuberger, Duffy, Brown, Morhishige,  Wallacher, 
Alba, Radhakrishnan} and \cite{Alba, Radhakrishnan, Dominguez, Han, Radhakrishnan2, Miyahara, 
Dijkstra, Hamada, Patrykiejew}, respectively. However, within the parameter range of the systems 
we are studying no genuine capillary freezing is expected to occur, because in accordance with 
the aforementioned studies in extended capillaries the freezing temperature is expected to be 
reduced compared with bulk freezing. An increase of the freezing temperature is expected to 
occur only for strongly attractive walls \cite{Radhakrishnan2, Miyahara} which is not the 
case here (see Sec. IV). Many details of capillary freezing also depend in addition on the 
capillary geometry (e.g. cylindrical versus slit pores \cite{Maddox}).

In Sec. II we provide a detailed description of the system studied here, which is characterized by
the texture of the surface as well as by the  fluid-fluid and fluid-surface interactions. In Sec. III 
we describe the theoretical technique we have used for the present investigation.  In Sec. IV we 
present and analyze our results.

\section{Wall topography and substrate potential}


\begin{figure*}[ht]

{\vspace{0.40cm}

\hspace{-12.0cm}\includegraphics[width=4.0cm, height=5.0cm]{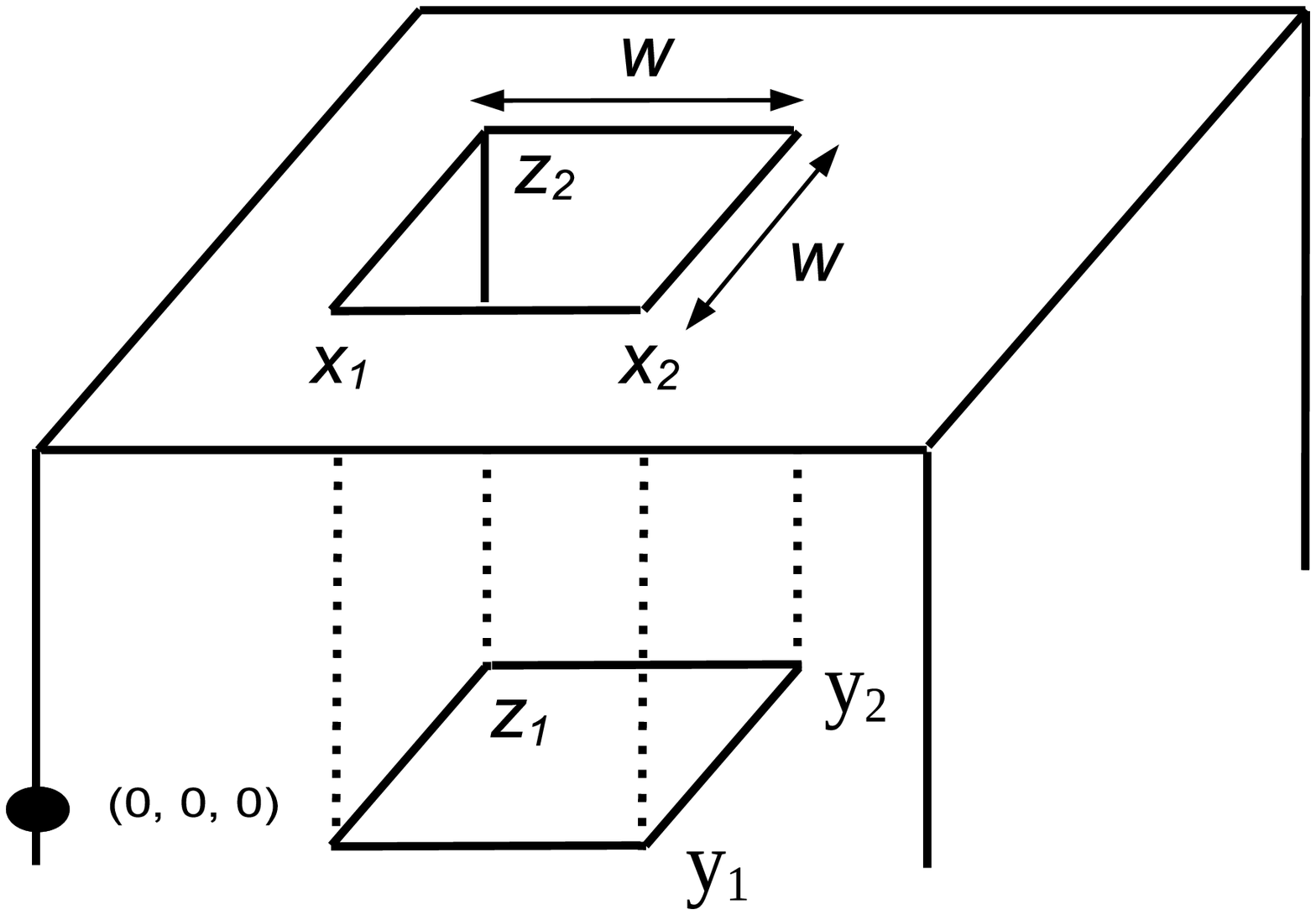}}

{\vspace{-2.40cm}
\hspace*{-5.2cm}\includegraphics[width=2.0cm, height=2.3cm]{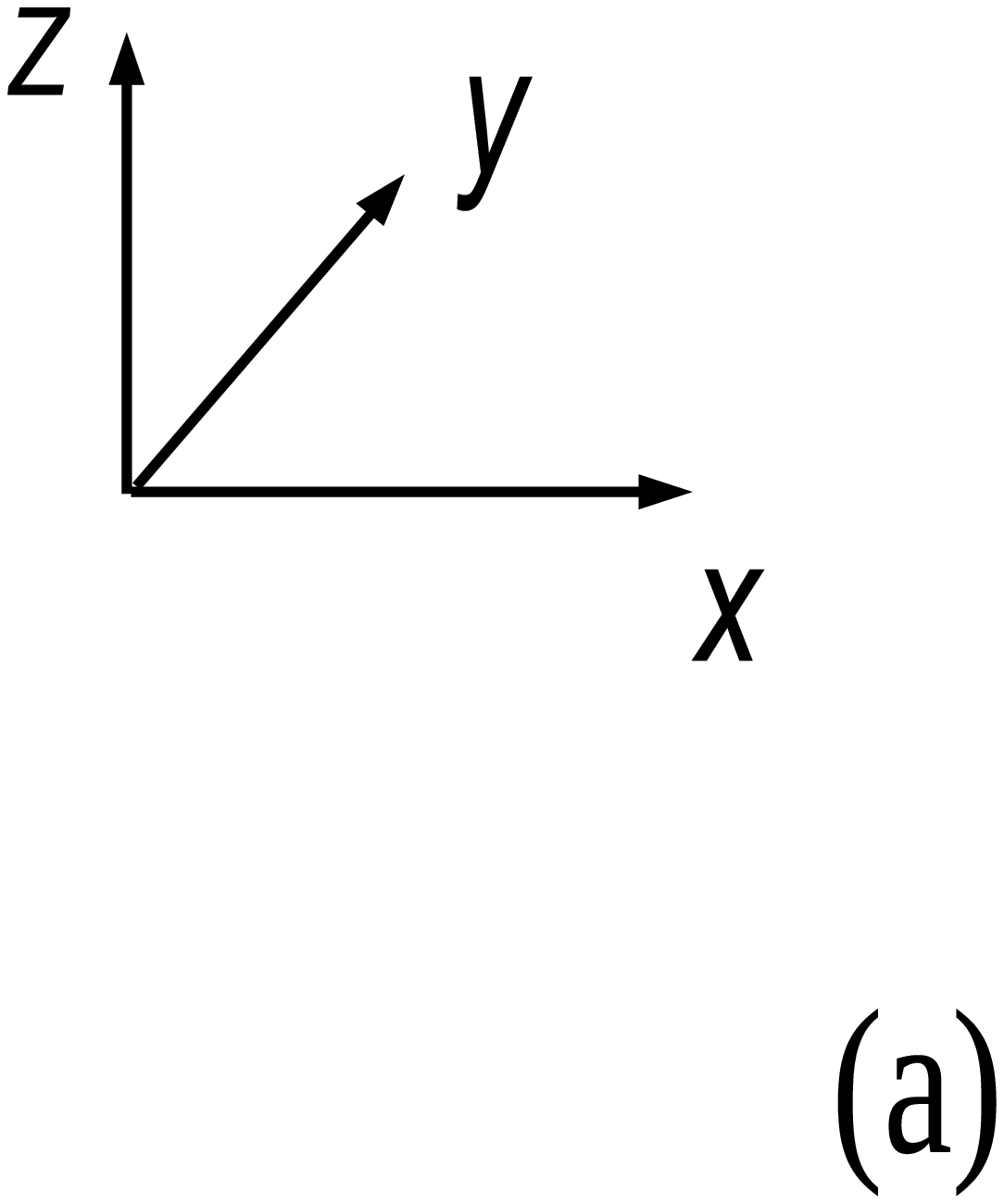}}

\vspace{-5.00cm}

{\hspace*{8.3cm}\includegraphics[width=7.5cm, height=5.0cm]{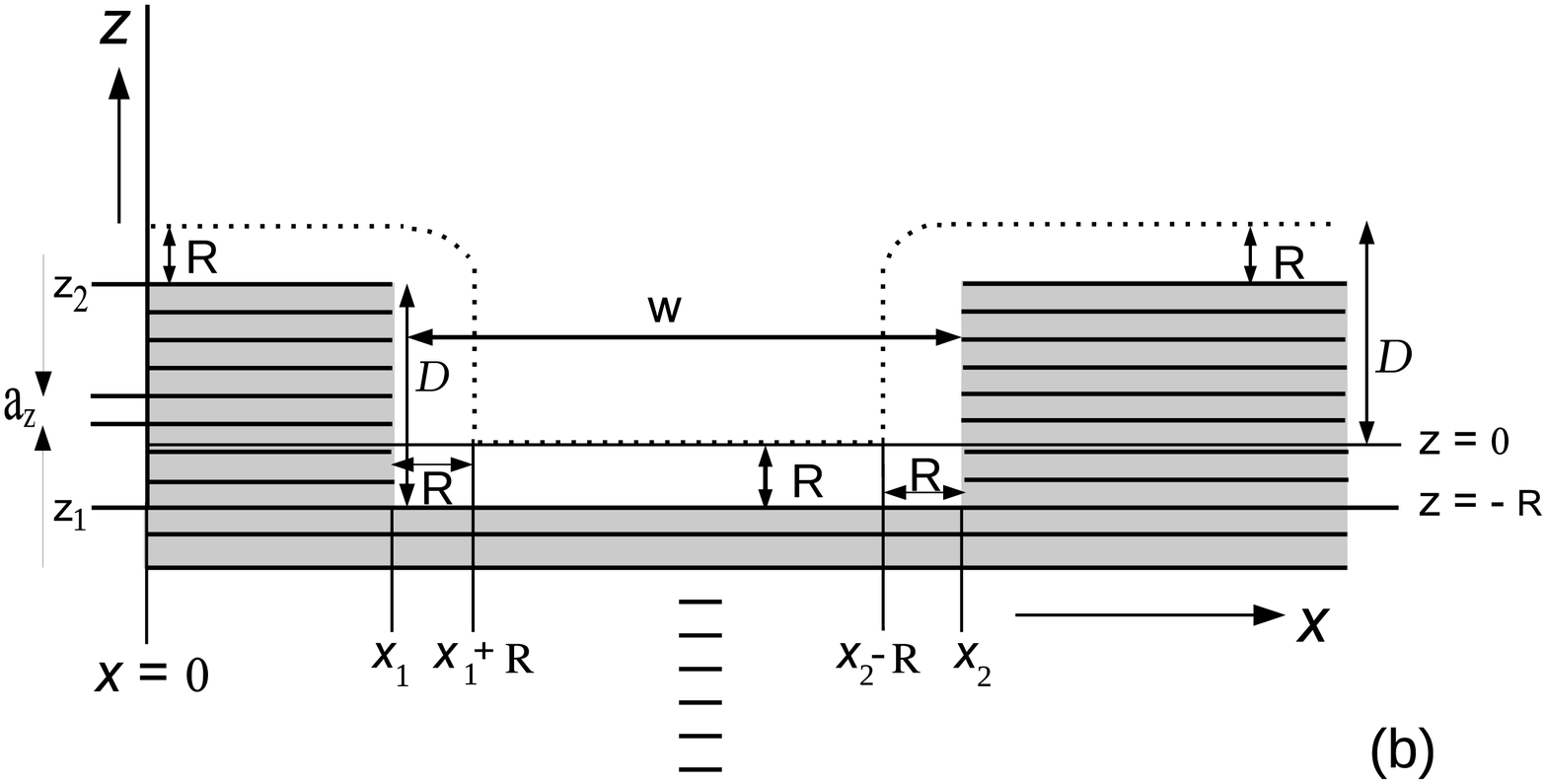}}

\caption{Schematic representation of a corrugated wall (a). The corrugation is modeled by square pits of 
width $w$ and depth $D$, which are independent of the arbitrary position of the origin. 
In (b) we show a vertical cut through the pit. The dotted lines indicate the depletion zone. 
Here $x_{1}=3.125 R$, whereas in the following figures we choose $x_{1} = R$. Note that $D$
is a multiple of $a_{z}$ but $R$ is not.} 
\end{figure*}

We have studied the structure of a fluid, which on one side is in its bulk liquid state and on 
the other side  in contact with a nano-structured wall. The wall exhibits a periodic array of 
square pits with edge length $w$ and depth $D$ as indicated in Fig . $1$. Depth and width of 
the pits are varied and the effect of the size of the pit on the 
fluid structure  is analyzed using density functional theory (DFT).  
The particles forming the liquid are considered to interact via a Lennard-Jones (LJ) pair potential 
{$\Phi_{LJ}(r)$}. 
In the spirit of DFT the free energy of the liquid is the sum of the free energy 
$F_{hs}$ of a reference system composed of hard spheres (hs) with pair potential

\begin{eqnarray}
 U_{hs}(r)&&=\infty,  \hspace{0.8cm} r\leq\sigma = 2R, \\ \nonumber
         &&= 0, \hspace{1.0cm} r > \sigma
\end{eqnarray}
and of a perturbative contribution $F_{att}$ to the free energy due to the attractive part $U_{att}(r)$
of the pair potential \cite{Weeks}:

\begin{eqnarray}
	U_{{att}}(r)&=&-\epsilon \Theta({2}^{1/6}\sigma-r) \\ \nonumber
			   &+&\Phi_{LJ}(r)\Theta(r-{2}^{1/6}\sigma) 
\end{eqnarray}
with the Heaviside function $\Theta$ and
\begin{equation}
\Phi_{LJ}=4\epsilon \left[{\left(\frac{\sigma}{r}\right)}^{12}-{\left(\frac{\sigma}{r}\right)}^{6}\right], 
\end{equation}
where $R$ is the radius of the fluid particles, $-\epsilon$ is the potential depth at $r=2^{1/6}\sigma$, 
$r$ is the center-to-center interparticle separation, and $\sigma$ is the distance of contact between 
the centers of two interacting liquid particles. The fluid-fluid interaction is rendered effectively 
short-ranged by introducing a cut-off at $r=R_{c}$ (in the following we adopt $R_{c}=5R$) 
which is implemented by a cut-off function which smoothly varies between $1$ at small 
distances and $0$ for  $r \geq R_{c}$. (The smoothness of the  cut-off improves the numerical stability
of our DFT calculations.) Close to $R_{c}$ this cut-off function drops from $1$ to $0$ continuously 
but very steeply and approaches $0$ at $R_{c}$ with zero slope. The interaction parameter $\epsilon$ for the cut-off 
potential is rescaled such that the integrated interaction is equal to the original one in Eq. $(2.3)$
with $R_{c} \to \infty$. 
This rescaling of the fluid-fluid interaction strength renders the liquid-vapor phase boundaries  
independent of $R_{c}$ within the DFT version used here, because for that version the bulk properties 
of the fluid depend only on the integrated interaction.

The repulsive part of the liquid-wall interaction is modeled by a hard core repulsion, so that 
the fluid particles cannot overlap with the wall, which is treated in this respect as a continuum 
with piecewise planar interfaces. This leads to a depletion layer of thickness $R$ above the wall
within which the fluid density is zero. The attractive part of the fluid-substrate (fs) interaction
is obtained by summing over the particles forming the wall:

\begin{equation}
U_{fs}(r)=-4\epsilon_{fs}\sum_{i}{\left[\frac{{\sigma_{fs}}^{2}}{{|{\bf r}-{{\bf r}_{i}}|}^{2}}\right]}^{3} ,
\end{equation}
where $\bf r$ is the position of a fluid particle, $i$ represents a wall particle at position 
${{\bf r}_{i}}$, $\epsilon_{fs}$ is the strength of the substrate potential, and $\sigma_{fs}$ 
is the distance of contact between a liquid and a wall particle.

We approximate the sum  in  Eq. $(2.4)$ as follows. We first calculate the contribution of 
a semi-infinite wall with its surface at $z=-R=-\sigma/2$ and extending down to $z=-\infty$ 
along the negative $z-$axis (see Fig. $1$). The wall is formed by layers with an interlayer spacing 
$a_{z}$. The layers are of macroscopic extent in  the lateral $(x, y)$ directions and treated as a 
continuum in the $x$ and $y$ directions. This leads to the following half-space contribution to the 
attractive fluid-solid interaction:


\begin{eqnarray}
	U^{half-space}_{fs}=&&- \biggl[\frac{u_{3}}{[(z+R)/R]^{3}} \\ \nonumber 
&&+\frac{u_{4}}{[(z+R)/R]^{4}} \\ \nonumber
&&+\frac{u_{5}}{[(z+R)/R]^{5}} +\cdots \biggr]  
\end{eqnarray}

Subsequently, we shall keep only the first three terms of this series; higher order terms are small for 
$z\geq0$ and $a_{z}/R \ll 1$ (e.g., with $a_{z}$ an atomic size and $R$ the radius of a colloid).
At closest contact the centers of the fluid particles are located at $z=0$.

The parameters in Eq. $(2.5)$ are related to the basic model parameters according to
$u_{3}=(\frac{2\pi}{3}){\epsilon'_{fs}}({\frac{\sigma_{fs}}{a_{z}}}) {(\frac{\sigma_{fs}}{R})}^{3}$,
$u_{4}=\frac{3}{2}(\frac{a_{z}}{R})u_{3}$, and $u_{5}={(\frac{a_{z}}{R})}^{2}u_{3}$, where ${{\epsilon'_{fs}}}$ 
is given by $\epsilon_{fs}$ times the number of wall particles within the area ${(\sigma_{fs})}^{2}$
in each of the layers building up the walls; if the wall particles forming a single layer occupy 
the sites of a simple $2d$ rectangular lattice with lattice constants $a_{x}$ and
$a_{y}$ this number equals ${(\sigma_{fs})}^{2}/(a_{x} a_{y})$ so that ${{\epsilon'_{fs}}}=\epsilon_{fs} 
{(\sigma_{fs})}^{2}/({a_{x} a_{y}})$.

On top of  this half-space we build the structured part of the wall by adding punctured layers 
({\it i.e.,} with missing squares of edge length $w$) which otherwise are continuous  and with 
vertical spacing $a_{z}$. They are added until the required depth $D$ is reached (see Fig. $1$). 
The interlayer spacing $a_{z}$ between these additional layers as well as the potential 
parameter ${{\epsilon'_{fs}}}$ characterizing the contribution of an area element of a layer
to the potential are chosen to be the same as for the layers forming the half space, {\it i.e.,} 
there is no chemical contrast between the half-space and the structured part of the wall built i
on top of it. Thus the total attractive part of the fluid-solid interaction is given by 

\be
U_{fs}({\bf r})=U_{fs}^{half-space}(z) + U_{structure}({\bf r})
\ee
with $U_{structure}({\bf r})$ determined as described above.

\section{Density Functional Theory}

The thermodynamic grand potential $\Omega$ of a classical one-component system characterized 
by positional degrees of freedom only is related to a functional of its number density $\rho({\bf r})$ 

\begin{equation}
\Omega[\rho]=F[\rho]+\int d^{3}r \rho({\bf r})(V_{ext}-\mu), 
\end{equation}
where $F[\rho]$ is the free energy functional, $V_{ext}({\bf r})$ an external potential, and $\mu$ 
the chemical potential. The equilibrium number density $\rho_{0}({\bf r})$ minimizes $\Omega$:

\begin{equation}
\frac{\delta \Omega[\rho]}{\delta \rho({\bf r})}|_{\rho(\br)=\rho_{0}(\br)}=0 .
\end{equation}
$\Omega[\rho_{0}(r)]$ is the grand potential of the system \cite{Evans2, gurug}.
The free energy functional can be decomposed as

\begin{equation}
F[\rho]=F_{id}[\rho]+F_{ex}[\rho]\nonumber
\end{equation}
with the ideal gas part  
\be
F_{id}[\rho]=k_{B} T \int d^{3} {r} \rho(\br)\left[\ln\left(\rho(\br)\Lambda\right)-1\right],
\ee
where $\Lambda={(\frac{h^{2}}{2\pi m k_{B}T})}^{3/2}$ is the cube of the thermal wave length 
associated with a particle of mass $m$; $h$ is the Planck constant and $k_{B}$ is the Boltzmann 
constant.

$F_{ex}$ arises due to interparticle interactions. It is the sum of 
two distinct contributions:

\begin{equation}
F_{ex}=F_{hs}+F_{att}
\end{equation}
where $F_{hs}$ corresponds to the contribution due to the hard core repulsion and is treated within 
the framework of fundamental measure theory (FMT), whereas $F_{att}$ arises due to the attractive 
part of the interaction and is treated within a simple random phase approximation.

\subsection{Fundamental Measure Theory}
The fundamental measure excess free energy for hard spheres as proposed by Rosenfeld is given by 
\cite{Roth3, Tarazona, Yasha1} 

\begin{equation}
\beta F_{hs}[\rho]=\int d^{3} r \hspace{0.1cm} {\phi(\{n_{\alpha}({\bf r})\})}
\end{equation}
where the excess free energy density $\phi$ is a function of weighted densities 
$n_{\alpha}({\bf r})$ defined as

\begin{equation}
n_{\alpha}({\bf r})=\int d^{3} r{^{'}}\rho({\bf r}^{'}) \omega_{\alpha}({\bf r}-{\bf r}^{'}), 
\end{equation}
where $\omega_{\alpha}$ are weight functions which characterize the geometry of the spherical particles.
These weight functions are obtained from the convolution decomposition of the volume excluded 
to the centers of a pair of particles in terms of the characteristic geometric features of the individual 
particles. They are given as 
\begin{eqnarray}
&&\hspace*{-1.0cm}\omega_{3}({\bf r})=\Theta(R-r), \hspace{0.1cm} \omega_{2}({\bf r})=\delta(R-r), \\ \nonumber 
&&\hspace*{-1.0cm}{\boldsymbol \omega}_{2}({\bf r})=\frac{\bf r}{r}\delta(R-r), \hspace{0.1cm} \omega_{1}({\bf r})=\frac{\omega_{2}({\bf r})}{4\pi R}, \\ \nonumber
&&\hspace*{-1.0cm}{\boldsymbol \omega}_{1}({\bf r})=\frac{{{\boldsymbol \omega}_{2}} ({\bf r})}{4\pi R}, \hspace{0.1cm} {\mbox{and}} \hspace{0.1cm} \omega_{0}({\bf r})=\frac{\omega_{2}({\bf r})}{4\pi R^{2}}, 
\end{eqnarray}
where $R$ is the radius of the spherical particles, $\Theta$ the Heaviside step function, and $\delta$ the 
Dirac delta function.

The original Rosenfeld functional fails to describe sharply peaked density distributions 
either in systems with an effective dimensionality smaller than $2$, such as liquids in extreme 
confinements, or in solids.
In order to overcome this problem, we have used a modified version of FMT, known as the modified 
Rosenfeld functional (MRF). The free energy density $\phi$ within the MRF framework is
\cite{Yasha2}  
\begin{eqnarray}
\phi&=&-n_{0}\ln(1-n_{3}) \\ \nonumber
&&+\frac{n_{1}n_{2}-{{{\bf n}_{1}} \cdot {{\bf n}_{2}}}}{1-n_{3}} \\ \nonumber
&&+\frac{({n_{2})}^{3}}{24\pi{(1-n_{3})}^{2}}{[1-{{\boldsymbol \xi}}^{2}]}^{q},
\end{eqnarray}
where $q\ge 2$ and ${\boldsymbol \xi}=\frac{{{\bf n}_{2}}}{n_{2}}$ (note that 
${{\bf n}_{2}} \cdot {{\bf n}_{2}} \neq {{(n_2)}^2}$). 
We have taken $q=3$ which reproduces
the original Rosenfeld functional up to the order ${\boldsymbol \xi}^{2}$.
For the contribution to the free energy due to the attractive part of the interaction the following 
truncation of the corresponding functional perturbation expansion is used:
\begin{equation}
F_{att}=\frac{1}{2}\int d^{3} r \int d^{3}r^{'}\rho({\bf r})\rho({\bf r^{'}})U_{att}({\bf r}-{\bf r^{'}}), \nonumber
\end{equation}
with $U_{att}$ defined in Eqs. $(2.2)$ and $(2.3)$. The minimization of $\Omega[\rho]$ 
has to be carried out numerically. To this end the
number density is discretized on a regular simple cubic grid. In order to obtain the results presented 
below we have used ${(\frac{R}{18})}^{3}$ as the elementary cube of the grid. We have carried out the Piccard 
iteration scheme to minimize $\Omega$ and to calculate the equilibrium number density. The weighted densities 
are calculated in Fourier space using the convolution theorem. More details about these techniques can be 
found in Ref. \cite{Roth4}. We have smeared out the distributions $\delta$ and $\Theta$ 
in order to achieve stable convergence for the convolution. The corresponding smearing 
length is a fraction of the distance between grid points and can be varied over a broad range without 
noticeably changing the results.
\section{Results}
In this section we present our results for equilibrium number density profiles of the liquid in 
contact with walls endowed with a periodic array of pits as described in Sec. II (see Fig. $(1)$).
The lateral size of the computation box is $14 R \times 14 R$ and the center to center distance between 
the pits is $14 R$. Sufficiently far above the wall ({\it i.e.,} in positive $z$ direction) the liquid 
becomes homogeneous and its density is given by the bulk density in the reservoir which is characterized 
by its packing fraction $\eta$. Accordingly, at the upper end of the computation box ({\it i.e.,} for 
$z > z_{max}$, with $z_{max}$ typically between $23R$ and $30R$) bulk boundary conditions are used: 
$\rho({\bf r})=\rho_{b},$ where $\rho_{b}$ is the number density in the bulk liquid with packing fraction

\be
\eta=\frac{4\pi R^{3}\rho}{3} \nonumber. \\
\ee
In the following number densities are given in units of $\frac {1}{R^{3}}$. In the transverse 
directions $x$ and $y$ periodic boundary conditions are used corresponding to the lateral periodicity 
of the array of pits.

In the following, we concentrate on the liquid structure inside the pits which is virtually
unaffected by the presence of the neighboring pits and therefore the precise size of the computational 
box  in the transverse directions is irrelevant in the present context.

All numerical results presented in the following have been obtained for a reduced temperature 
$T^{*}=\frac{k_{B}T}{\epsilon}\equiv 1$, where $\epsilon$ corresponds to the strength of the non-truncated
attractive fluid-fluid interaction $(R_{c}\to \infty)$. (For comparison, the model exhibits the critical
temperature ${T_{c}}^*$ $\approx$ $1.43$ and the triple point temperature ${T_3}^{*}\approx 0.692$ as obtained via
MC simulations \cite{Barroso}.) For ${T^*} = 1$ the fluid model
we are using renders $\eta_{l}^{(lv)}$ $\approx$ $0.346$ for the liquid packing fraction at bulk liquid-vapor
coexistence; the corresponding vapor packing fraction is $\eta_{v}^{(lv)}$  $\approx$  $0.010$. 
Concerning the packing fractions at  liquid-solid coexistence we rely on simulation results for Lennard-Jones liquids 
\cite{Barroso} which give the packing fractions $\eta_{l}^{(ls)}=0.483$  and $\eta_{s}^{(ls)}=0.528$ for the 
coexisting liquid and solid, respectively, at $T^{*} = 1$.
Our numerical results correspond to bulk packing fractions between $0.38$ and $0.46$, which 
are sufficiently apart from both the liquid-vapor and liquid-solid phase boundaries so that
neither drying nor capillary freezing occurs.

For the parameters characterizing the fluid-wall interaction (Eq. $(2.5)$) we have used
$u_{3}=1.0$, $u_{4}=0.0375$, and $u_{5}=0.000625$, respectively (the chosen ratios $u_{4}/u_{3}$ and
$u_{5}/u_{3}$ correspond to an interlayer distance $a_{z}/R=1/40$), in units of $k_{B}T = \epsilon$.  
For these parameters, according to Young's equation at $T^{*}=1.0$ 
the contact angle formed by the coexisting liquid and vapor phases at a planar wall is about $160^{o}$.
For packing fractions above ca. $0.40$, in addition to vertical layering, ``{\it three-dimensional} " 
localization of the liquid within the pits may occur, {\it i.e., } spots may form within which the 
liquid density is high and which are separated from each other in all spatial directions by regions in 
which the density of liquid is considerably lower. We have varied the width $w$ and the depth $D$ of the pits 
and have studied how the fluid structure within the pits evolves as a function of these geometric parameters.
We mainly present results obtained for $\eta=0.42$; the liquid structure corresponding to higher  
values of $\eta$ are qualitatively similar to the ones corresponding to $\eta = 0.42$. 
In the following all lengths are given in units of $R$.  

\subsection{Structure of high density regions as function of the width of the pits}

\begin{figure*}[ht]
\hspace{-0.8cm}\includegraphics[height=1.4in,width=1.3in]{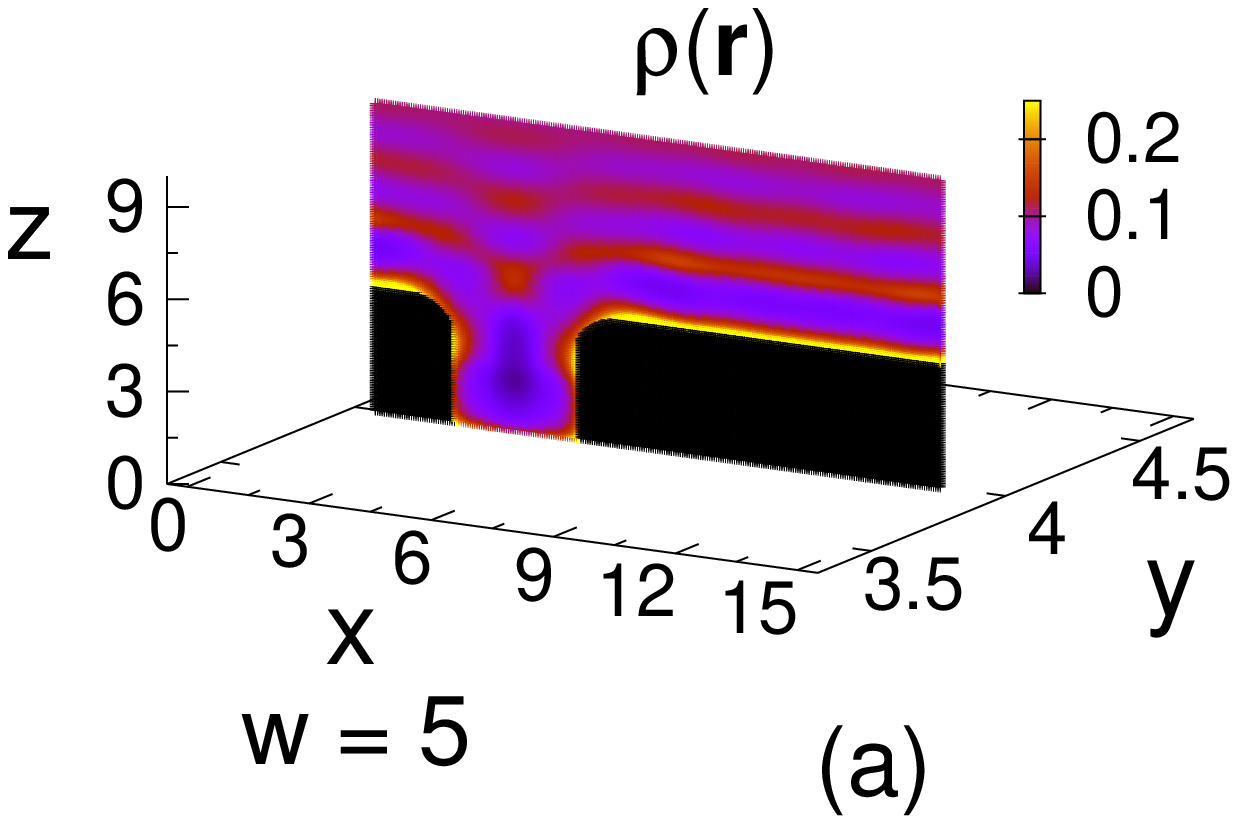}
\hspace{0.41cm}\includegraphics[height=1.4in,width=1.3in]{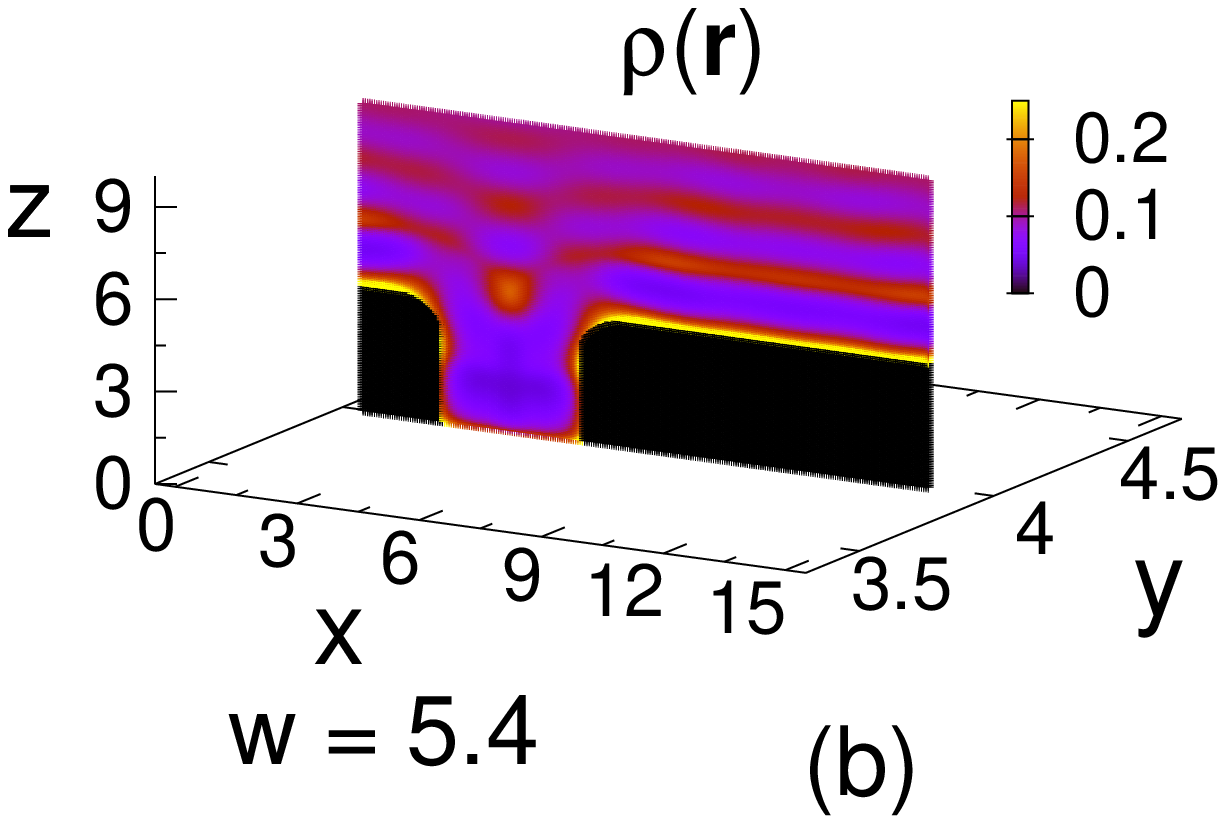}
\hspace{0.41cm}\includegraphics[height=1.4in,width=1.3in]{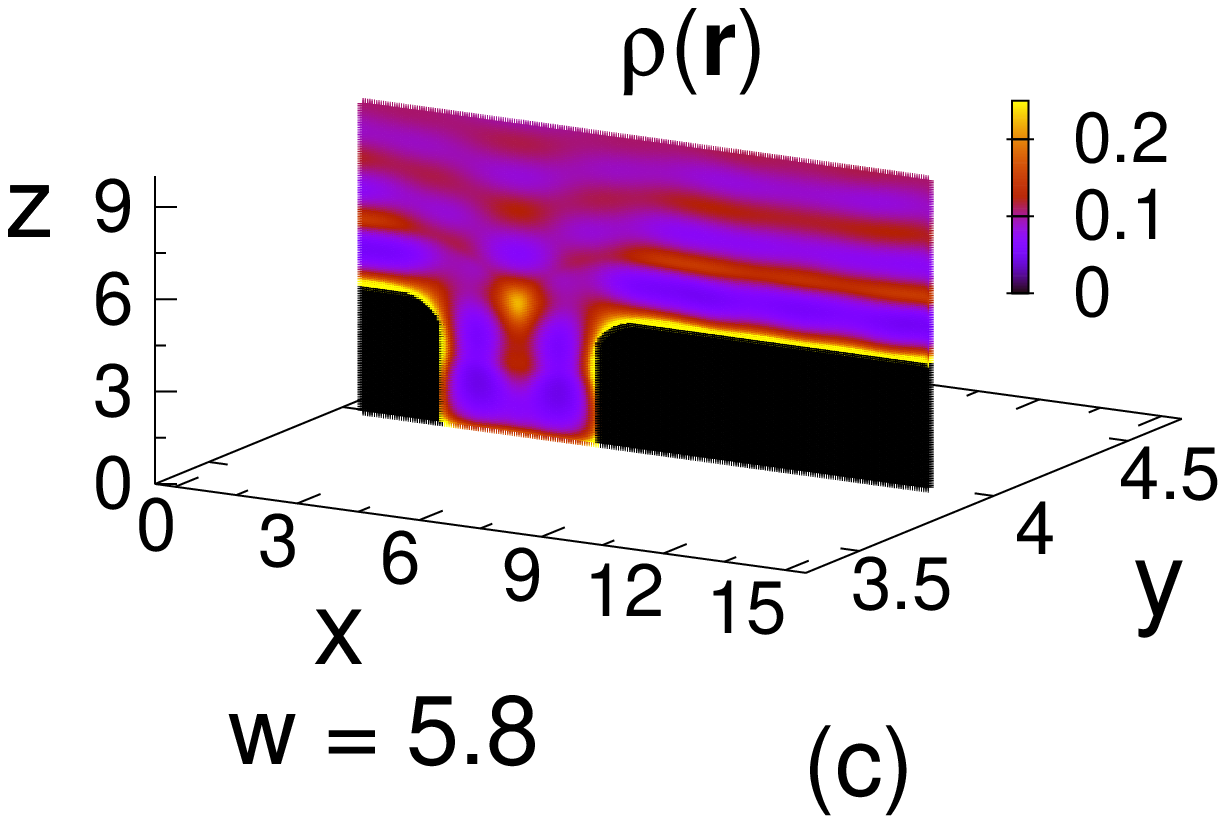}
\hspace{0.41cm}\includegraphics[height=1.4in,width=1.3in]{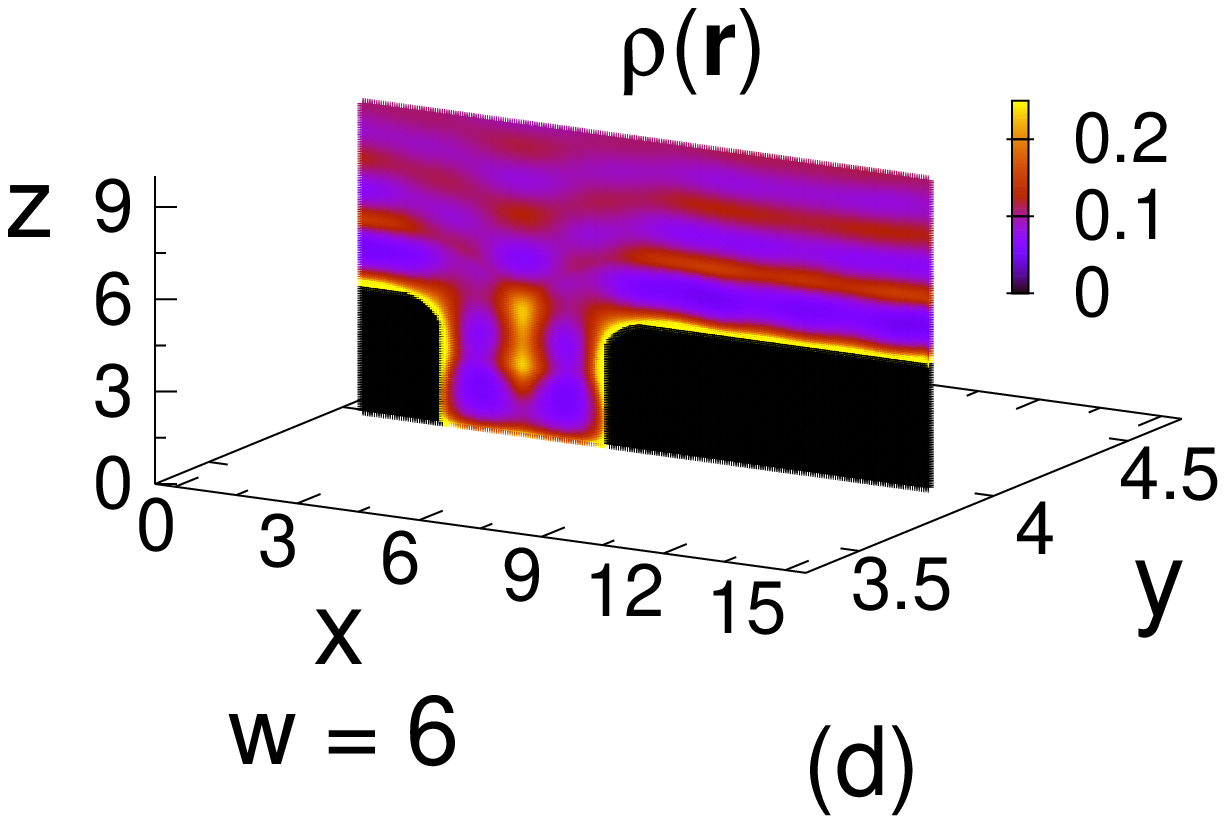}

\vspace{0.5cm}

\hspace{-0.8cm}\includegraphics[height=1.6in,width=2.3in]{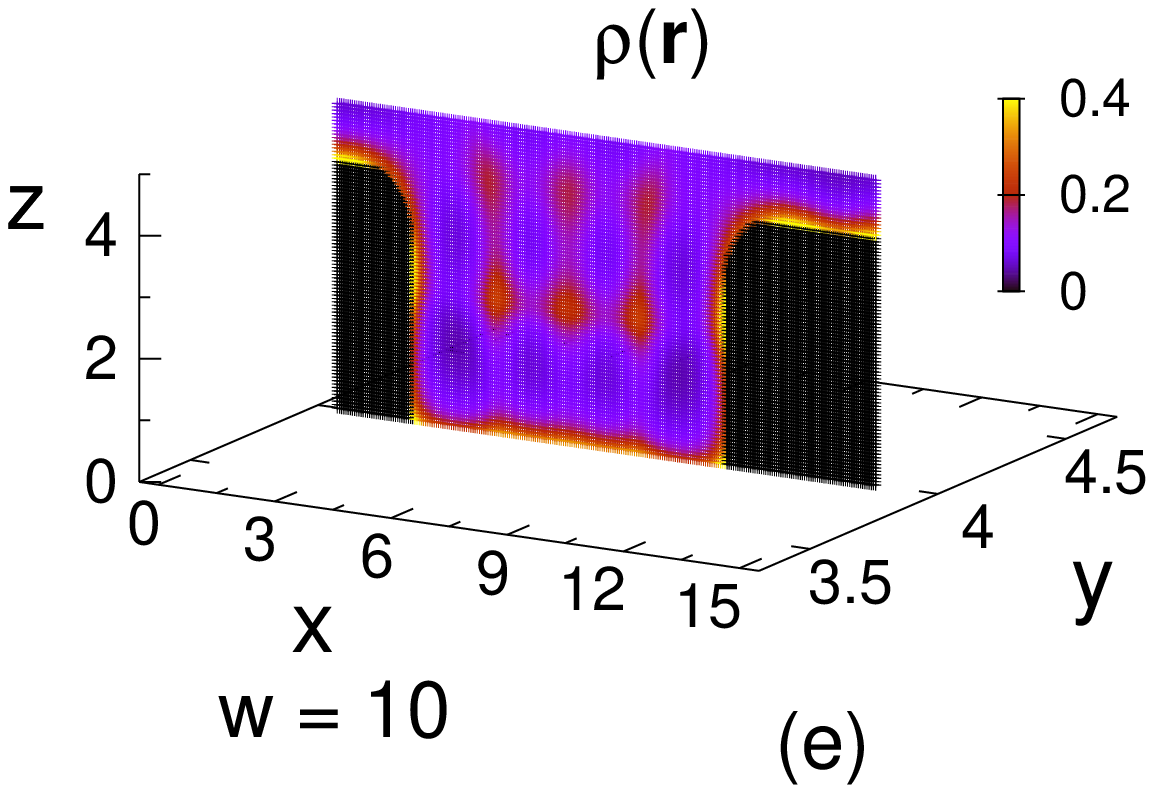}
\hspace{2.0cm}\includegraphics[height=1.6in,width=2.3in]{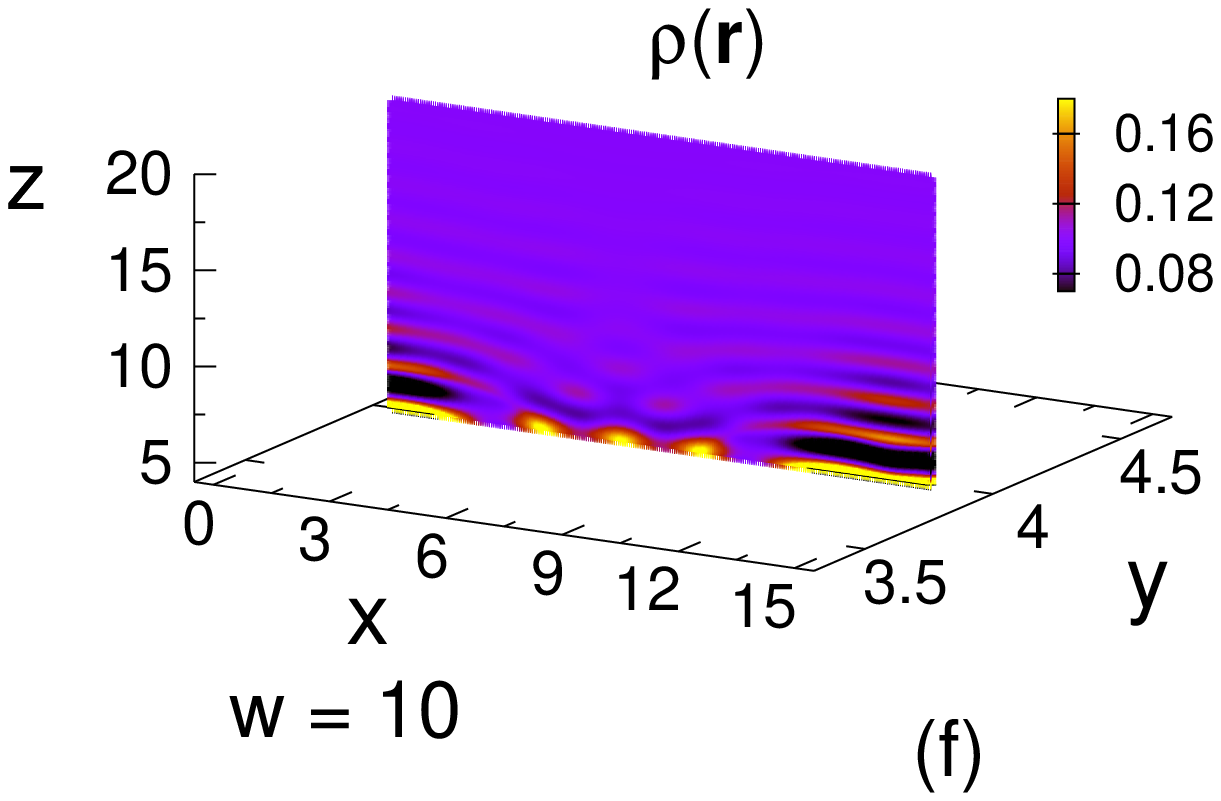}

\vspace{-0.2cm}
\caption{The number density $\rho(\br)$ in the $xz$ plane $y=4.0$ for $\eta=0.42$
(which corresponds to $\rho_{bulk}= 0.10$), $D=4.0$, and for various widths $w$ ((a)-(d)). The
origin $(x, y, z) = (0, 0, 0)$ of the coordinate system is chosen such that the left corner
at the front and bottom of the pit is located at $(x, y, z) = (x_{1}, y_{1}, z_{1}) = (1, 1, -1)$
(see Fig. $1$). The positions $x_{2}$ and $y_{2}$ are varied in order to implement various pit widths.
For $w=10.0$, (e) and (f) compare the densities inside and above the pit. Note that the color coding
between (e) and (f) as well as between (a)-(d) and (e)-(f) is different. The excluded volume is indicated
in black; $z<0$ is excluded. }
\end{figure*}

In Fig. $2$ we show the liquid densities at cross sections through pits of various widths $w$.
The depth $D$ of the pit is kept constant. The cutting plane is a $xz$ plane, perpendicular to 
the substrate surface and parallel to the front side wall of the pit chosen to be at a distance 
$3R$ from that side wall. By that it passes through the second layer of enhanced densities parallel
to that front side wall and exhibits a density distribution typical for the analogous layers observed 
in the interior of the pit. The first layer, {\it {i.e.}}, the one just next to the depletion zone is 
somewhat special due to its overall enhanced density and is therefore less suited for characterizing 
the liquid structure. Indeed, it is visible that the density is somewhat enhanced at the walls, 
{\it i.e.,} next to the excluded volume. For geometric reasons there is some rounding of the exclusion 
zone at the upper edges of the pit.  Accordingly, the pit is effectively slightly wider at its upper end 
than at its bottom. This explains why a spot of high liquid density appears at the pit opening first 
as the width of the pits increased to $w = 5.4$ (Figs. $2(a)$ and $(b)$). The density at this spot increases 
further upon widening the pit (Fig. $2(c)$). For even wider pits the region of enhanced density spreads into 
the pit (Fig. $2(d)$).

\vspace{ 0.3cm}

\begin{figure*}[ht]


\hspace{-0.5cm}\includegraphics[height=2.1in,width=2.7in]{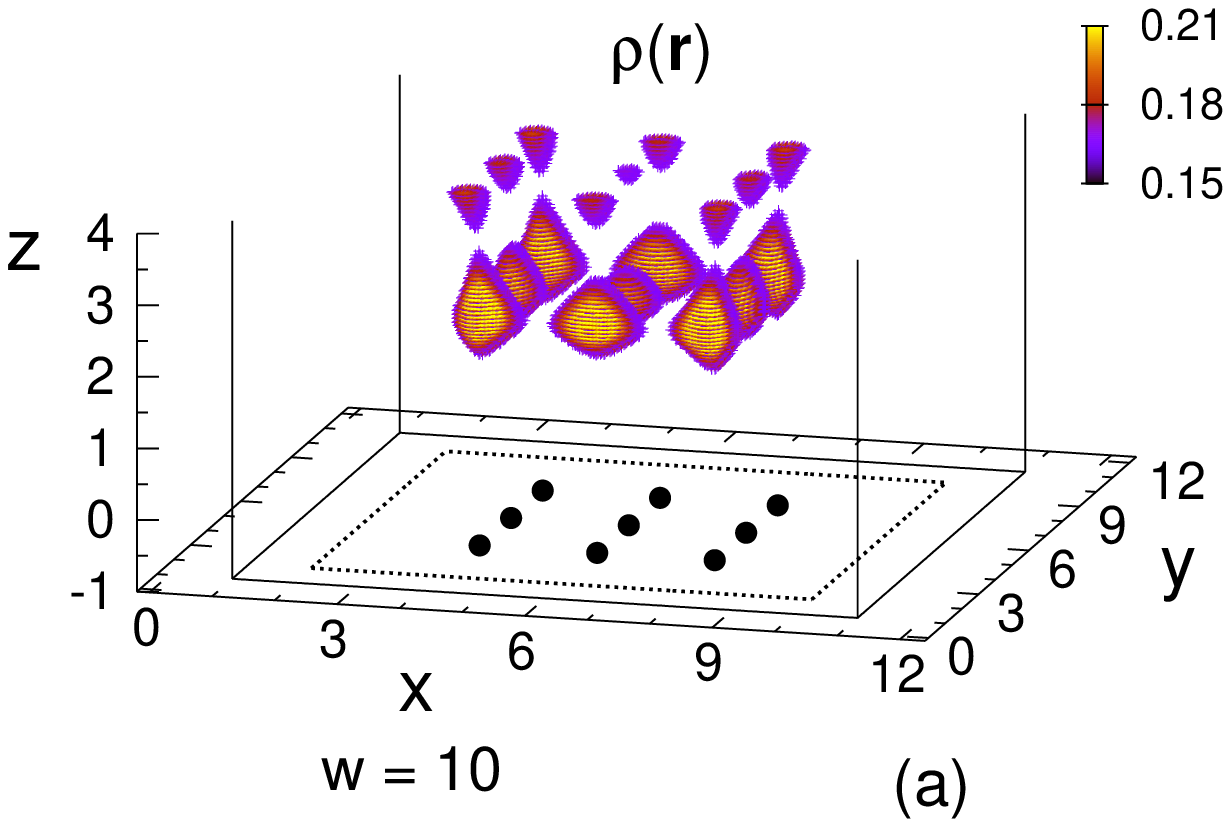}
\hspace{1.5cm}\includegraphics[height=2.1in,width=2.7in]{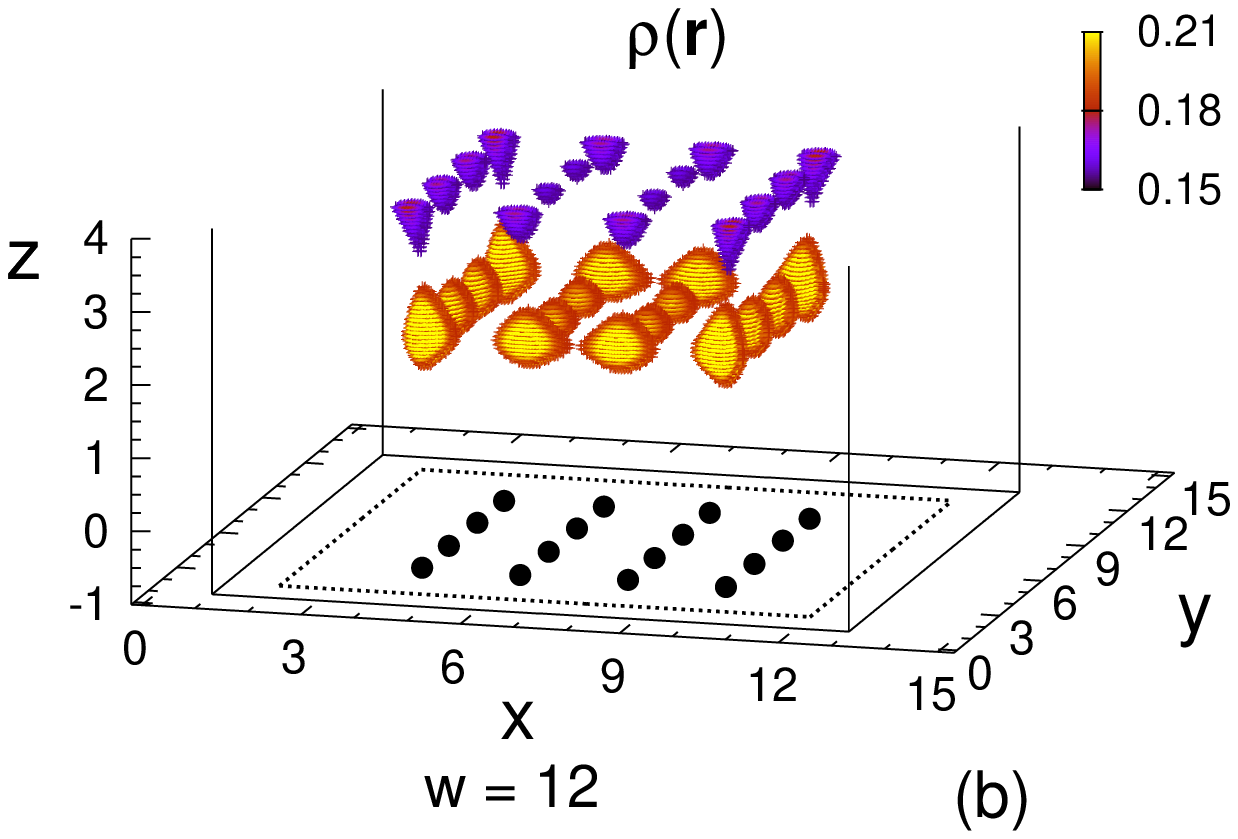}

\vspace{-0.4cm}

\caption{Structure inside the pit (high density regions right at the walls next to the depletion
zone are excluded) for $\eta=0.42$, $D=4.0$, and $w=10.0$ (a) and $w = 12.0$ (b). The left
bottom front corner of the pit is fixed at $(x, y, z) = (1, 1, -1)$ whereas the other corners are
shifted in order to change the width of the pit.  Only the regions with the density higher
than $0.17$ for the lower ({\it i.e.}, first) layer and regions with the density
higher than $0.15$ for the upper ({\it i.e.}, second) layer are shown in order to focus on the high
density regions only. Thin lines indicate the position of the walls forming the pit. The black
dots are projections of the centers of the spots of the lower layer onto the plane
$z = -1$. The projections of the boundaries of the depletion zone onto the plane $z = -1$ are indicated
by dotted black lines.}
\end{figure*}

Upon increasing $w$ further, additional structures emerge inside the pits. Upon reaching
$w=10.0$ the density increases both at the side walls of the pit and at its bottom 
(Fig. $2(e)$). Moreover, Fig. $2 (e)$ shows that three spots of high liquid density appear
in the first layer above the bottom layer and another three with somewhat weaker density 
in the second layer near the pit opening. This second layer does not develop in pits with $D < 4$. 
Outside the pits only layering is observed (see Fig. $2(f)$).  Above 
the pit opening the layers are bent downwards. However, within a few particle diameters 
these distortions die out both in the vertical and in the lateral directions.

\begin{figure*}[ht]

\vspace{0.7cm}

\hspace{-2.3cm}\includegraphics[height=1.4in,width=2.2in]{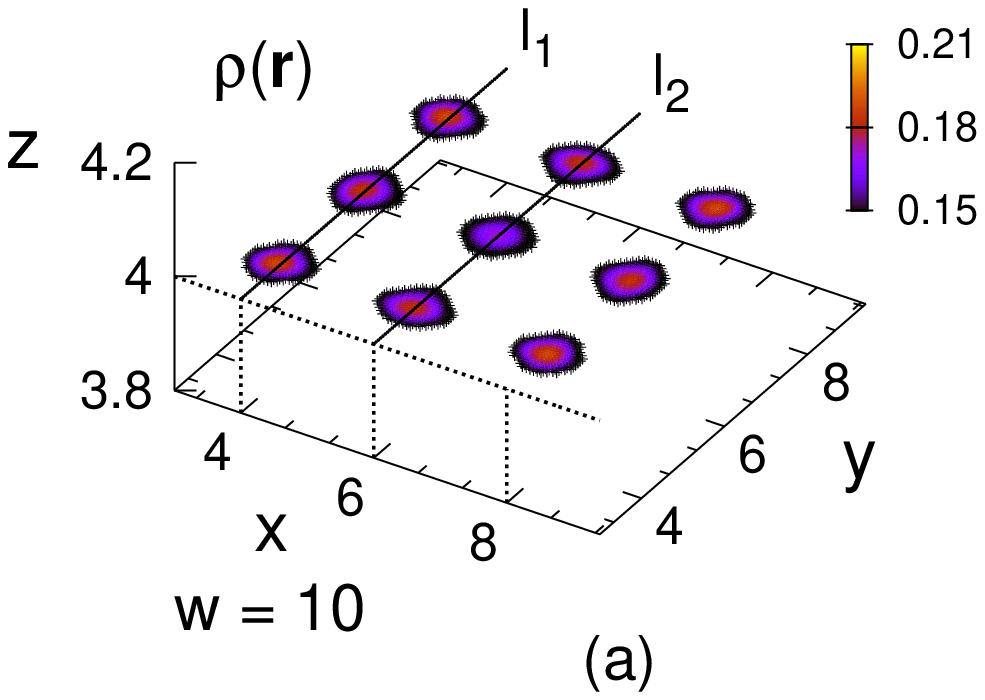}
\hspace{2.4cm}\includegraphics[height=1.4in,width=2.2in]{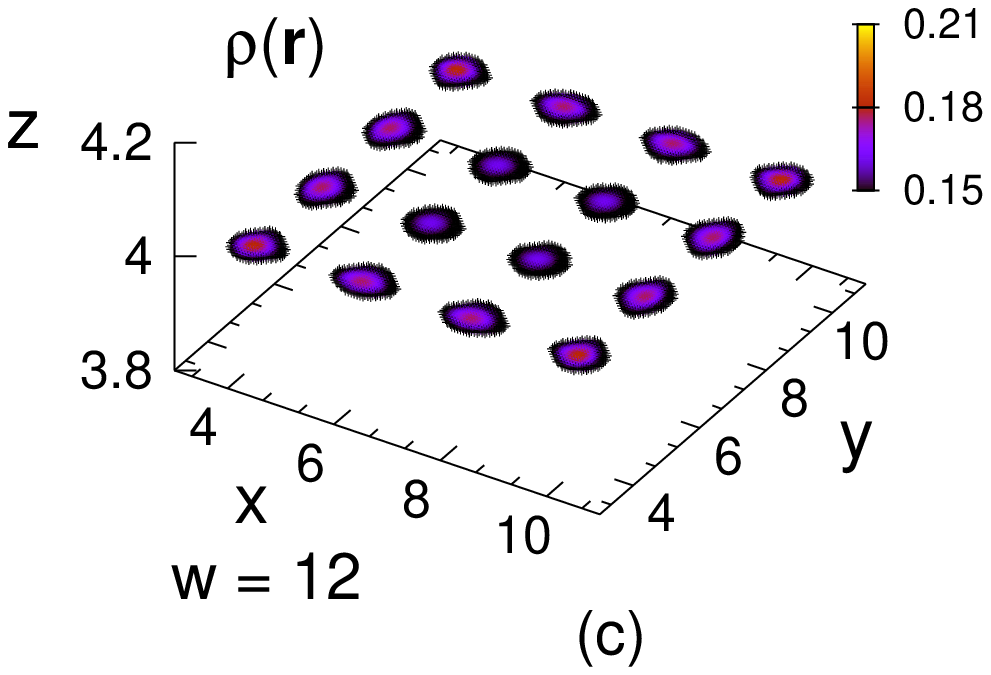}

\vspace{0.6cm}

\hspace{-2.3cm}\includegraphics[height=1.4in,width=2.2in]{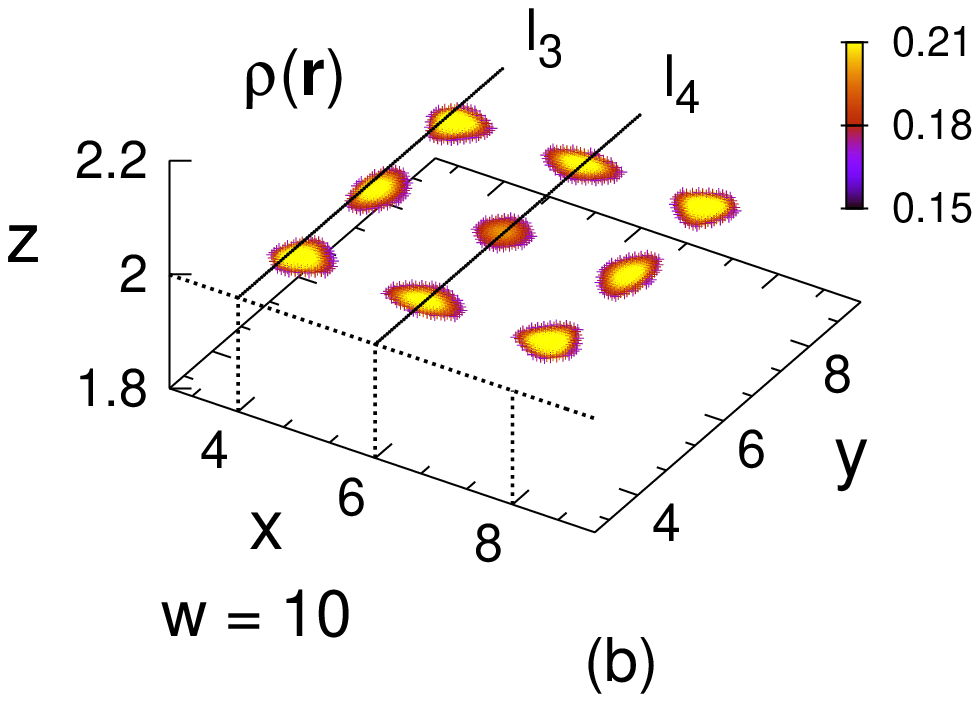}
\hspace{2.4cm}\includegraphics[height=1.4in,width=2.2in]{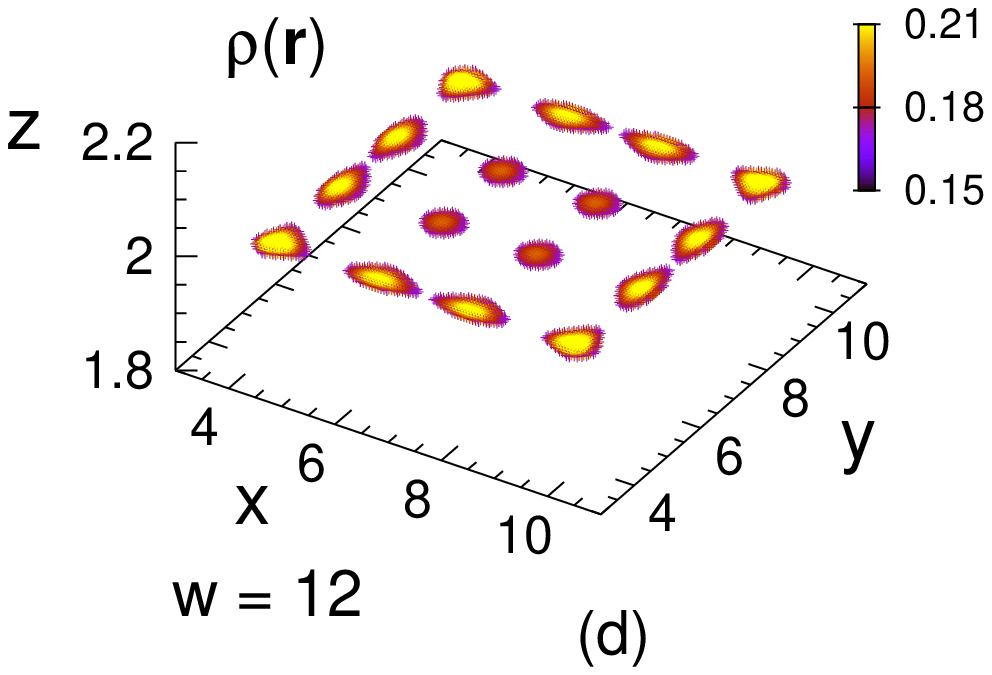}

\vspace{-0.2cm}

\caption{Horizontal cuts through the upper ((a), (c)) and the lower ((b), (d)) layer
for the structures shown in Fig. $3$. The density thresholds are $0.15$ in (a) and (c) and $0.17$ in
(b) and (d). For various values of $w$, the density distributions along the lines $l_{1}$ and $l_{2}$ in the
upper layer (a) are shown in Figs. $7(a) - 7(f)$ and Figs. $7(g) - (j)$, respectively. The density
distributions along the lines $l_{3}$ and $l_{4}$ in the lower layer ((b)) are shown in Figs.
$8(a) - (f)$ and Figs. $8(g) - (j)$, respectively. The lines $l_{1} - l_{4}$
run through the center of the spots shown in (a) and (b).}
\end{figure*}

Two alternative representations of the liquid structure (for $\eta=0.42$ and $w=10.0$) within the pits are 
shown in Fig. $3(a)$ and Figs. $4(a)$ and $4(b)$. In Fig. $3$, we show the three-dimensional configuration 
of the high density spots in the interior of the pit, {\it i.e.}, here we do not consider the layers of 
high liquid density adjacent to the inner walls of the pit. Furthermore, we show only the densities 
exceeding certain threshold values in order to render the configuration of the high density spots visible. 
According to Fig. $3(a)$, inside the pit two horizontal layers emerge, each of which forming a $3 \times 3$
array of high density spots. This can be seen even more clearly in Figs. $4(a)$ and $4(b)$ which 
shows horizontal cuts through both the second and the first layer, again only displaying densities above a 
certain threshold value. The threshold value is set to roughly $1.5$ times the density averaged  
over the inside of the pit where the density is nonzero. For $\eta=0.42$ this averaged density is $0.1144$ and $0.1133$ 
for $w=10$ and $w=12$, respectively,  and the threshold value for the densities displayed in 
Figs. $3$ and $4$ is taken to be $0.17$ ($\approx 1.5 \times 0.1133$). For comparison, in 
Figs. $3(b)$, $4(c)$, and $4(d)$ we show the same kind of representations but for wider 
pits ($w=12.0$, $D=4.0$). In this case, again two layers emerge, but each forming a  $4 \times 4$
array of high density spots. Figure $5$ illustrates the pathway leading from a $3 \times 3$ to 
a $4 \times 4$ structure upon increasing the pit width gradually from $w=10.0$ to $w=12.0$.

\begin{figure*}[ht]
\hspace{-0.8cm}\includegraphics[height=1.2in,width=1.5in]{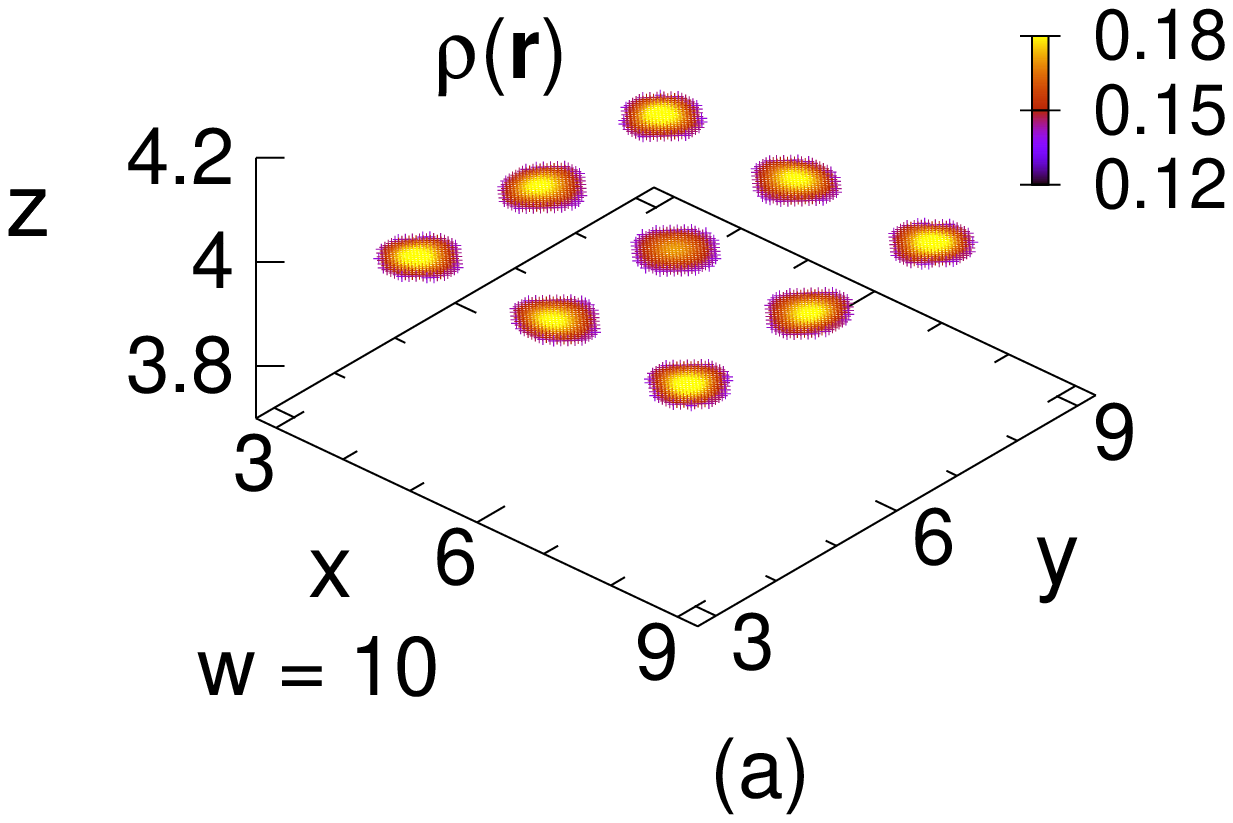}
\hspace{0.1cm}\includegraphics[height=1.2in,width=1.5in]{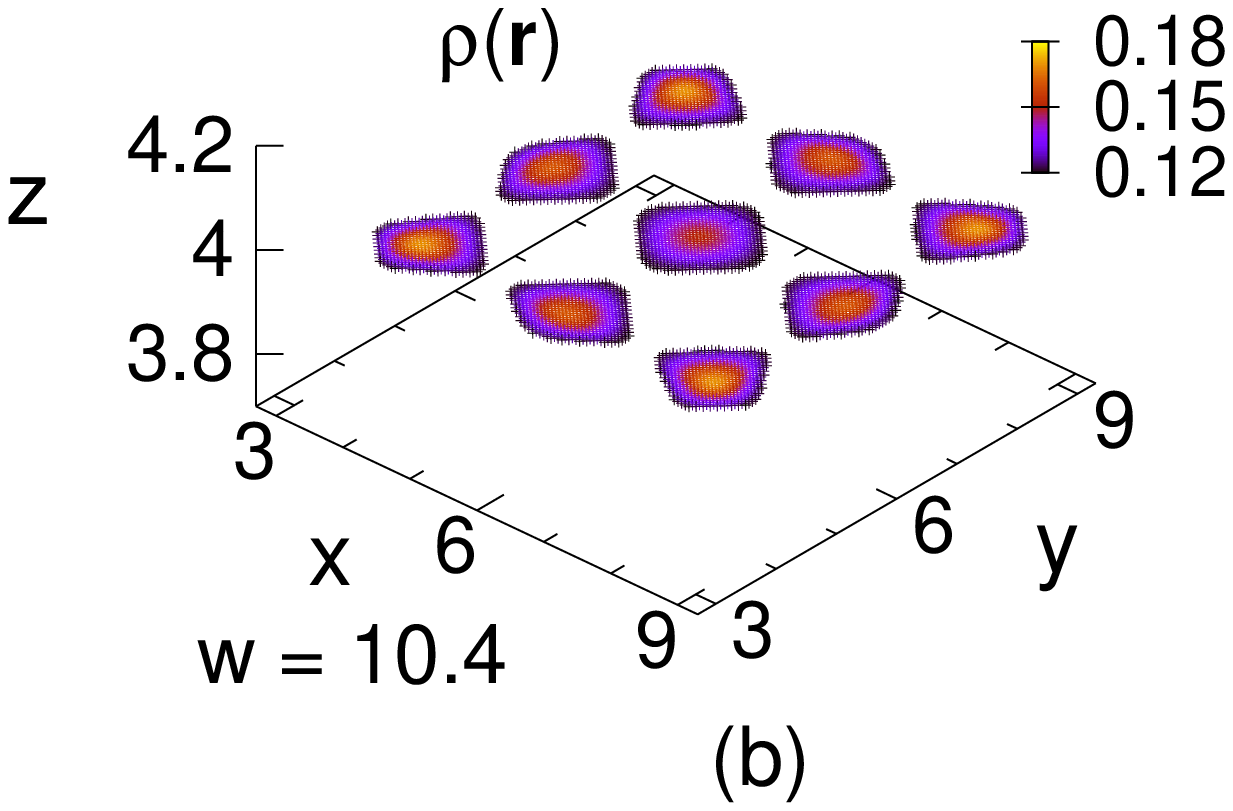}
\hspace{1.1cm}\includegraphics[height=1.2in,width=1.5in]{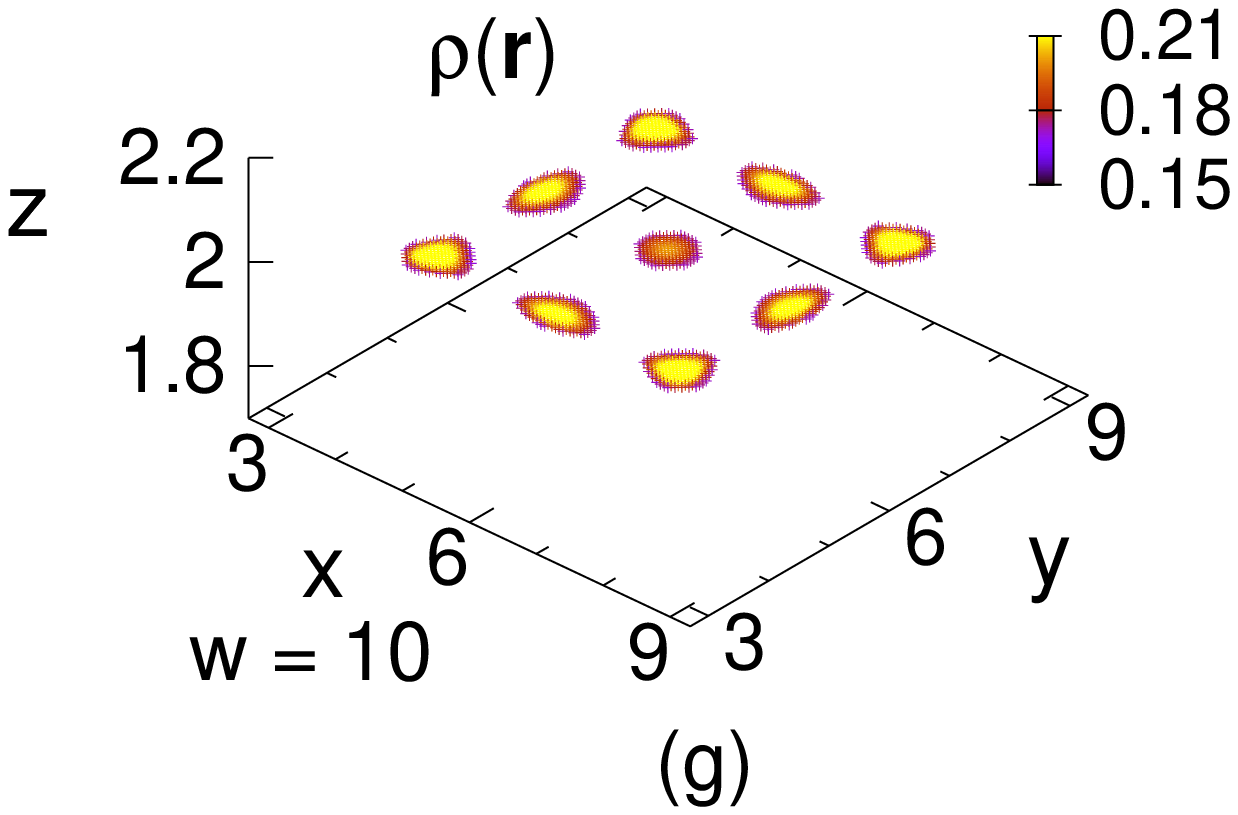}
\hspace{0.1cm}\includegraphics[height=1.2in,width=1.5in]{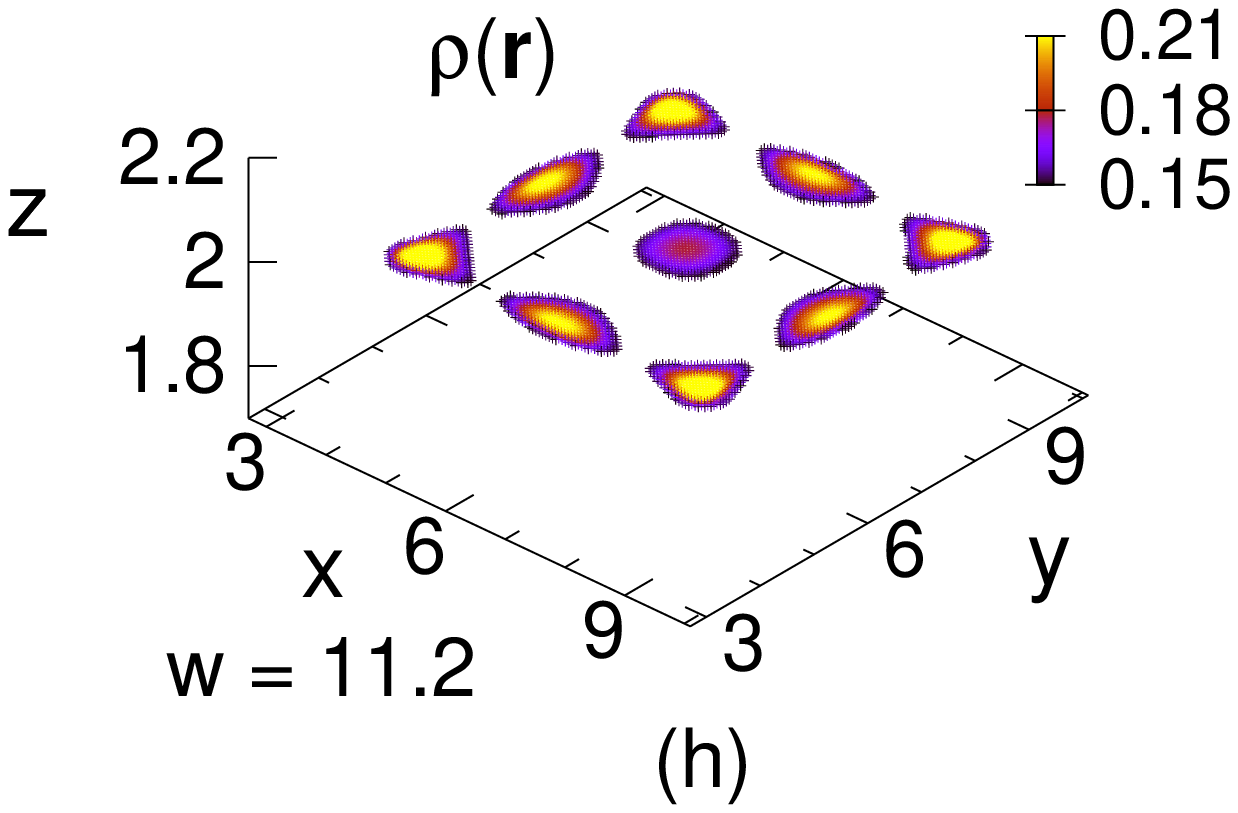}

\vspace{0.4cm}
\hspace{-0.8cm}\includegraphics[height=1.2in,width=1.5in]{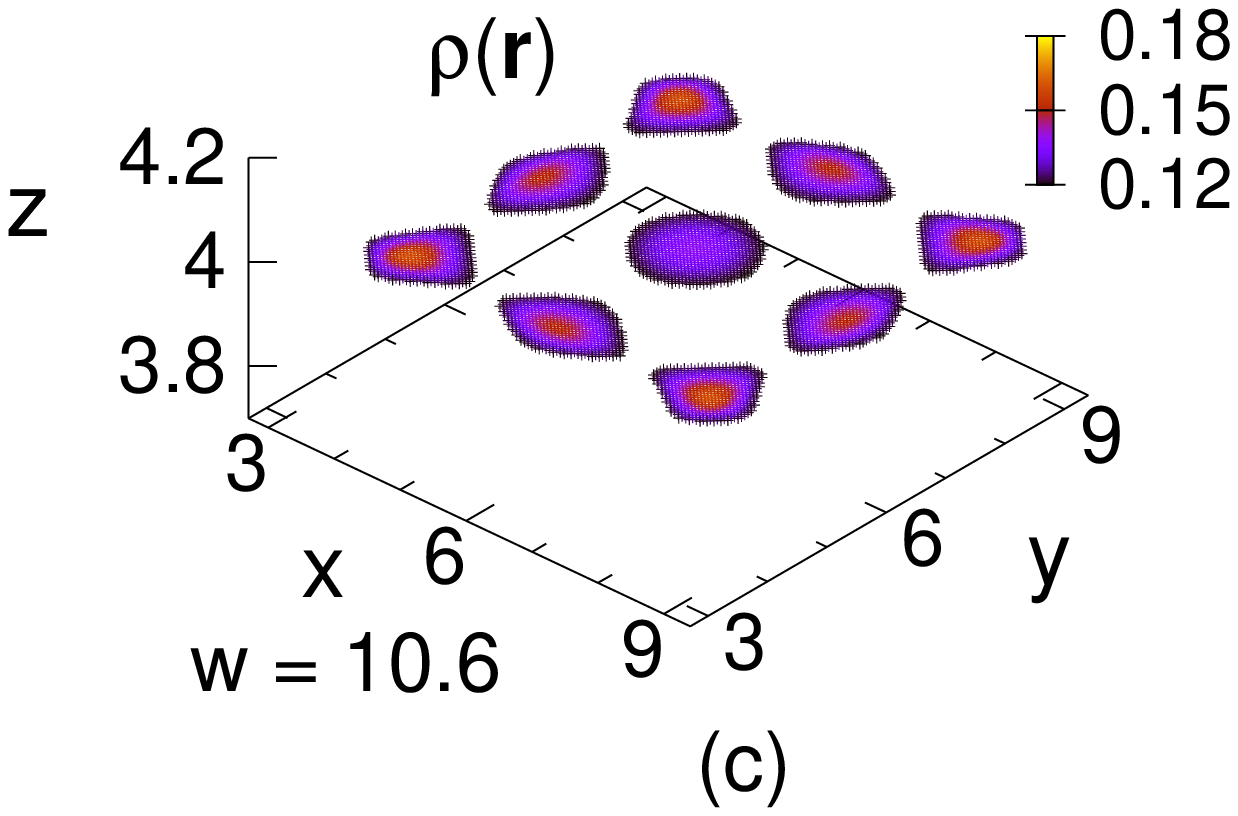}
\hspace{0.1cm}\includegraphics[height=1.2in,width=1.5in]{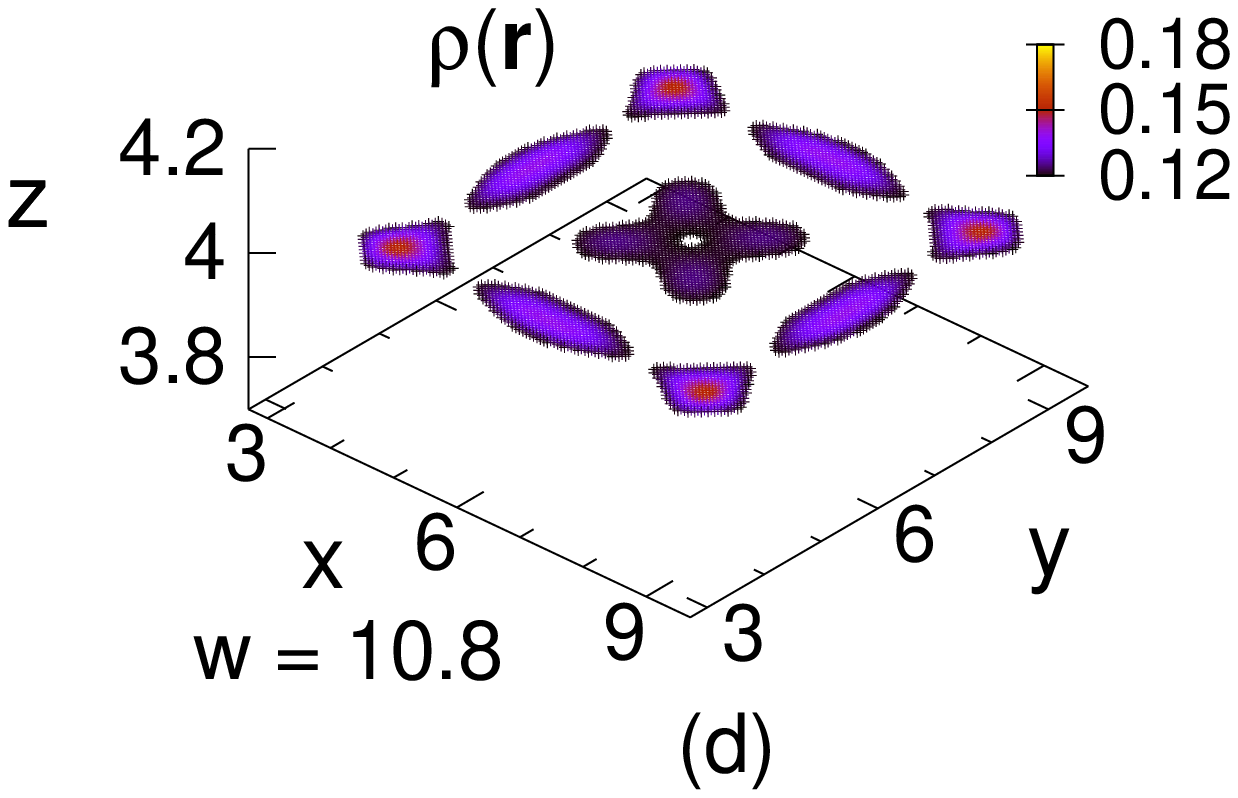}
\hspace{1.1cm}\includegraphics[height=1.2in,width=1.5in]{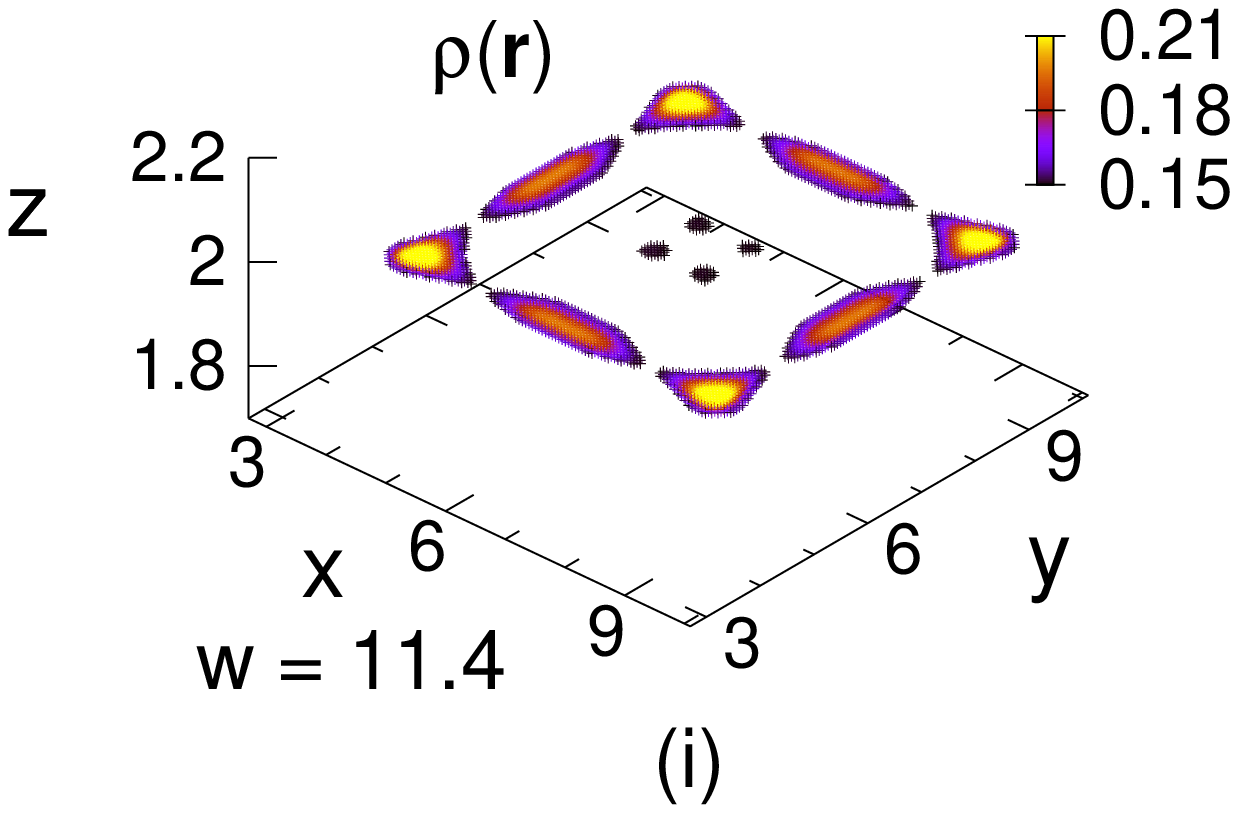}
\hspace{0.1cm}\includegraphics[height=1.2in,width=1.5in]{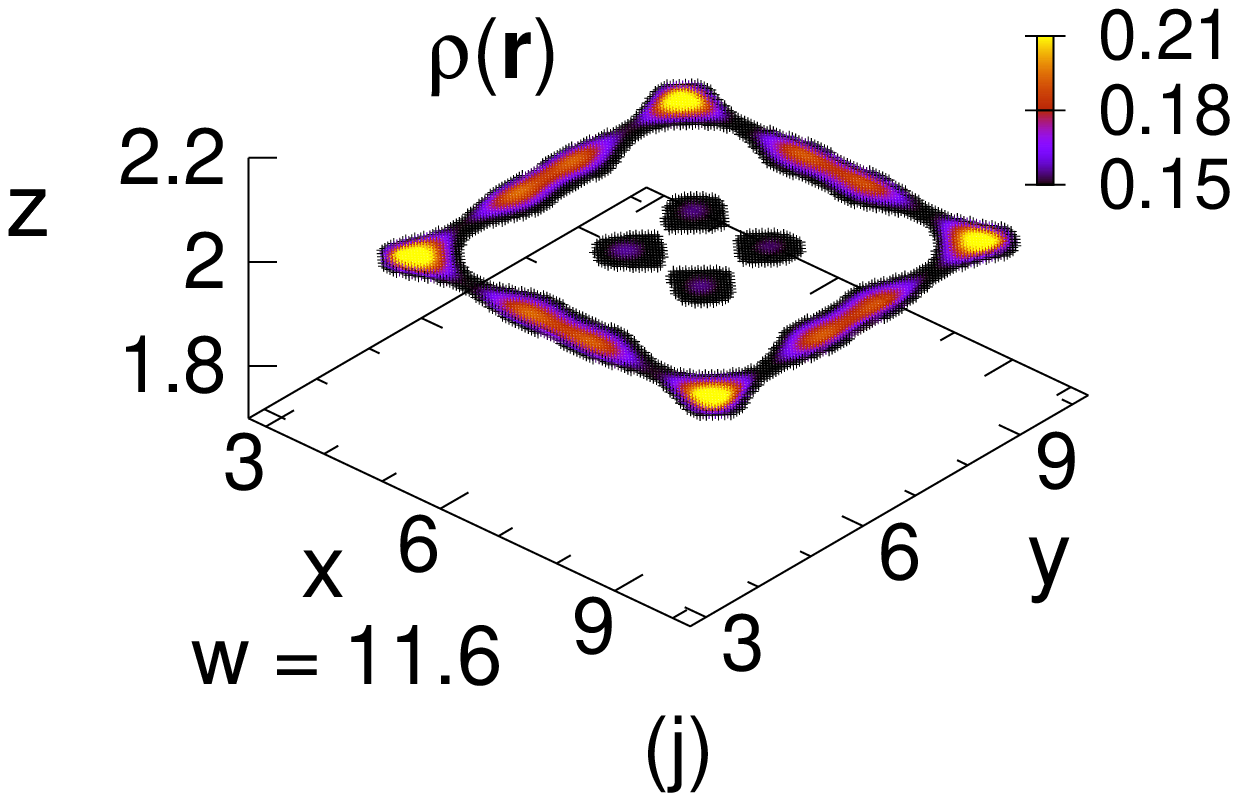}

\vspace{0.4cm}

\hspace{-0.8cm}\includegraphics[height=1.2in,width=1.5in]{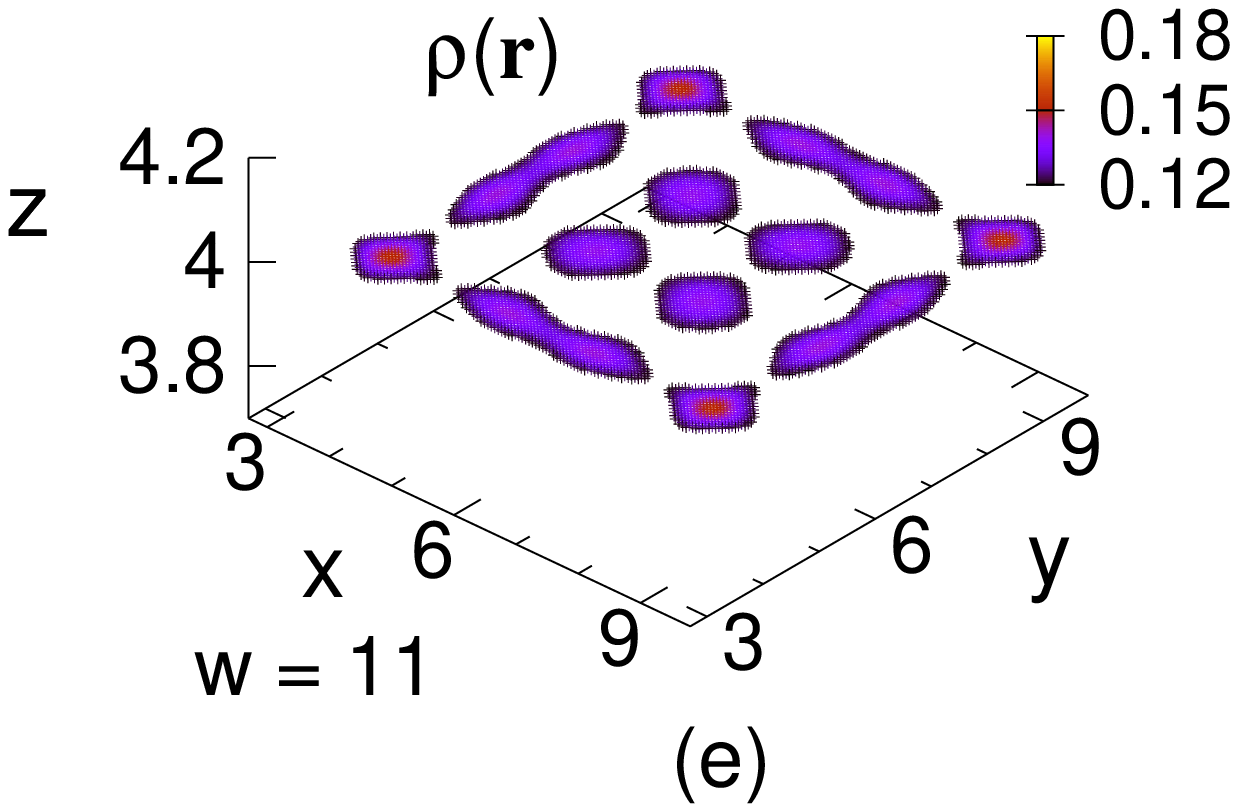}
\hspace{0.1cm}\includegraphics[height=1.2in,width=1.5in]{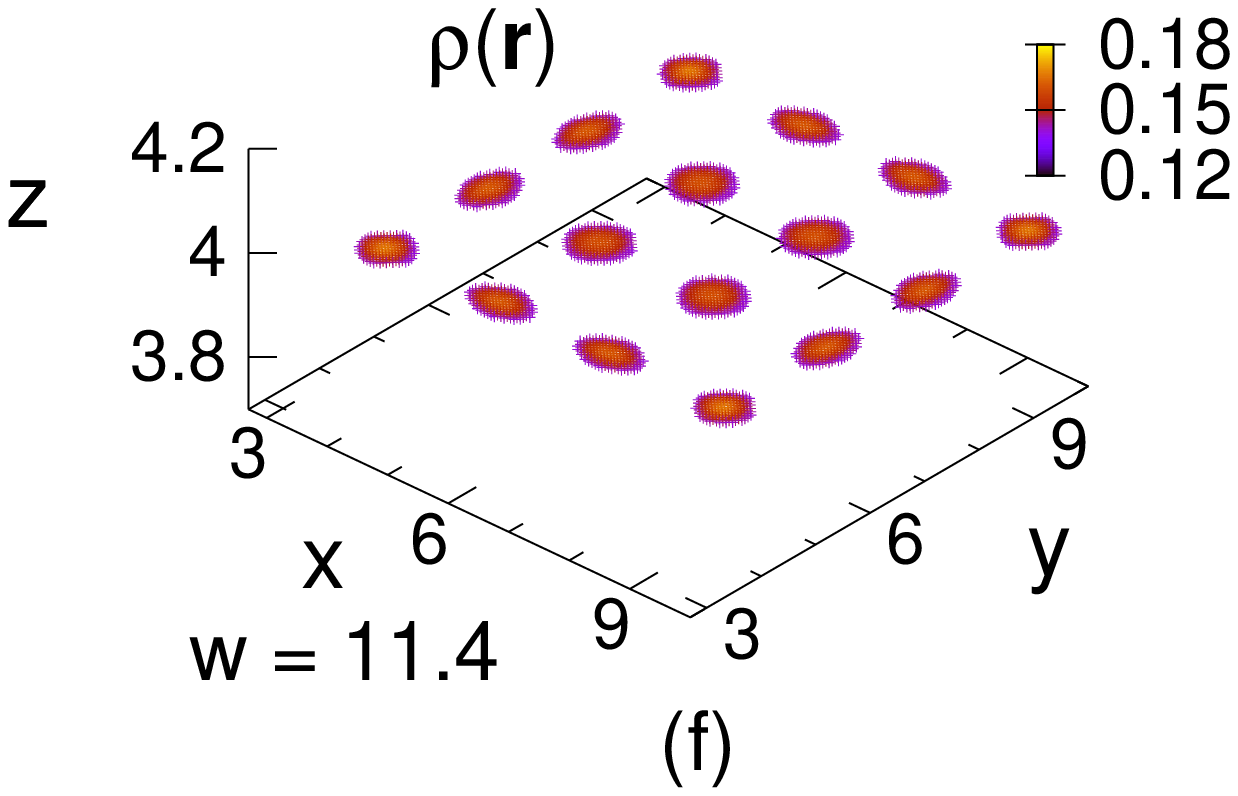}
\hspace{1.1cm}\includegraphics[height=1.2in,width=1.5in]{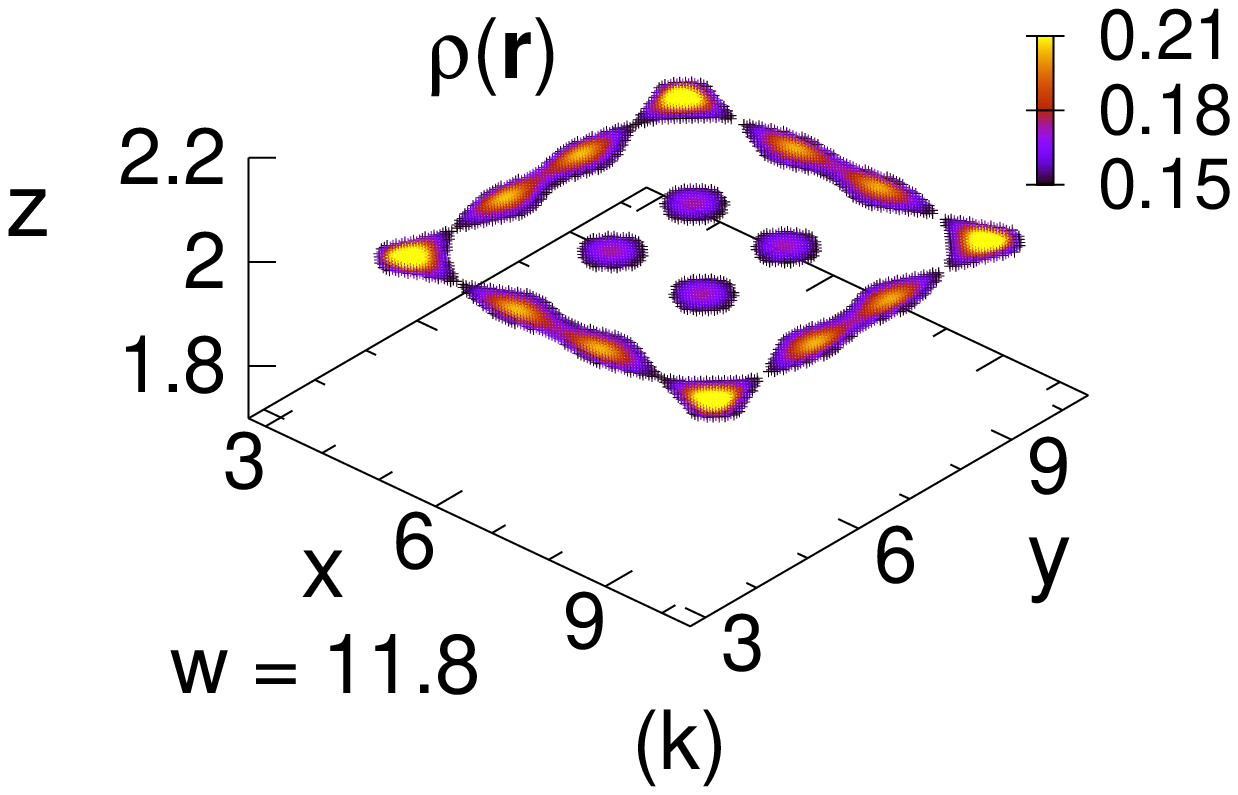}
\hspace{0.1cm}\includegraphics[height=1.2in,width=1.5in]{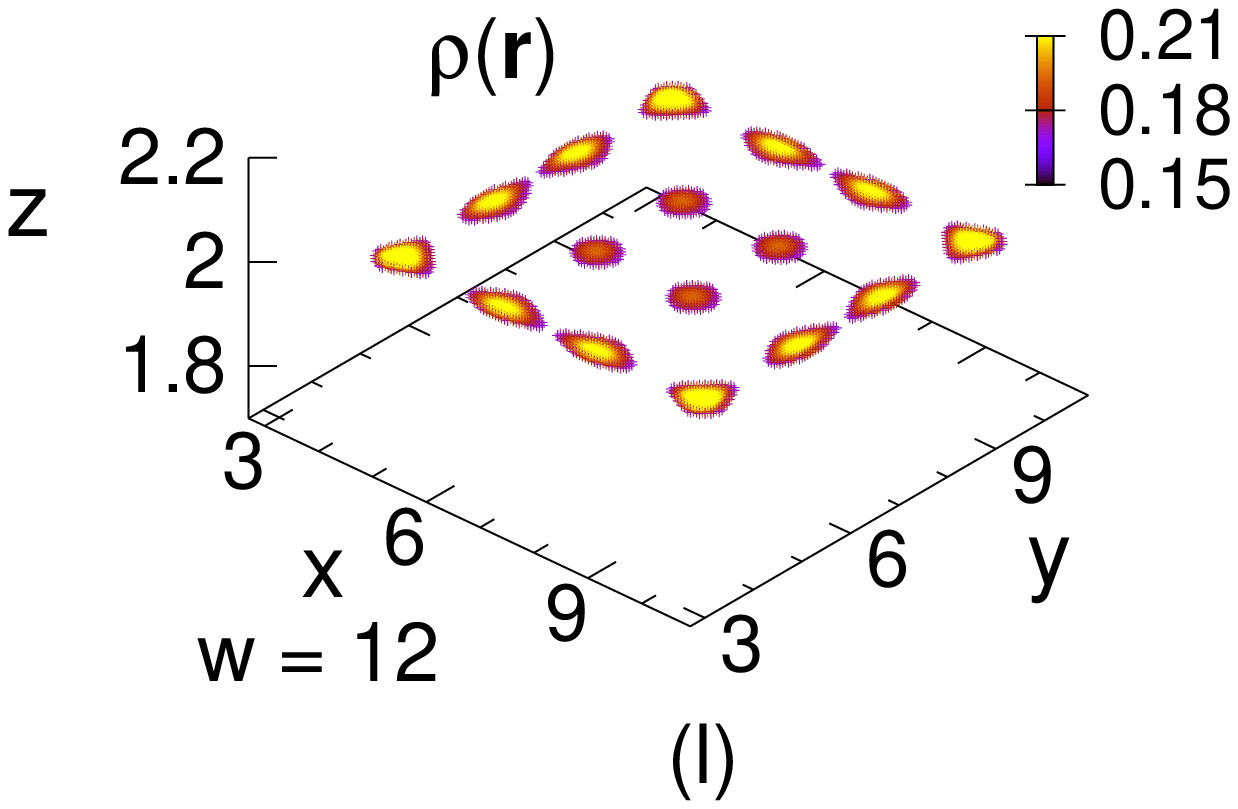}

\vspace{-0.2cm}

\caption{Density distribution $\rho(\br)$ in the plane $z=4$ (upper ({\it i.e.,} second) layer)
((a) - (f)) and in the plane $z=2$ (lower ({\it i.e., } first) layer) ((g)-(l)) for $\eta=0.42$,
$D=4$, and for various values of $w$. Only the regions with density above $0.12$ are shown for the
upper layer ((a)-(f)) and regions with the density above $0.15$ are shown for the lower layer ((g)-(l)).
Here the threshold is lower than in Figs. $3$ and $4$ because the average density inside
the pits, to which the threshold is proportional, is smaller for the incommensurate pit widths (see Fig. $9$ ).}

\end{figure*}

\begin{figure*}

\vspace{0.4cm}

\hspace{-0.2cm}\includegraphics[height=1.6in,width=1.3in]{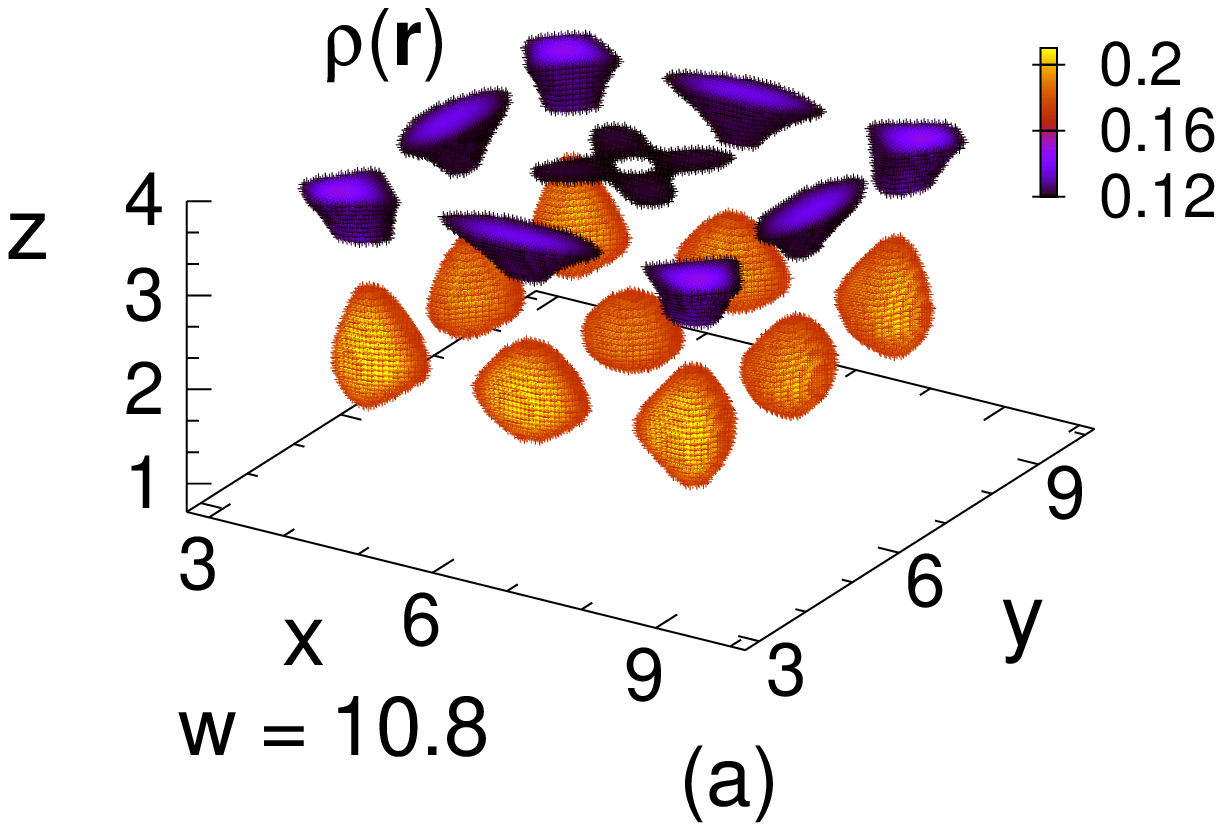}
\hspace{0.6cm}\includegraphics[height=1.6in,width=1.3in]{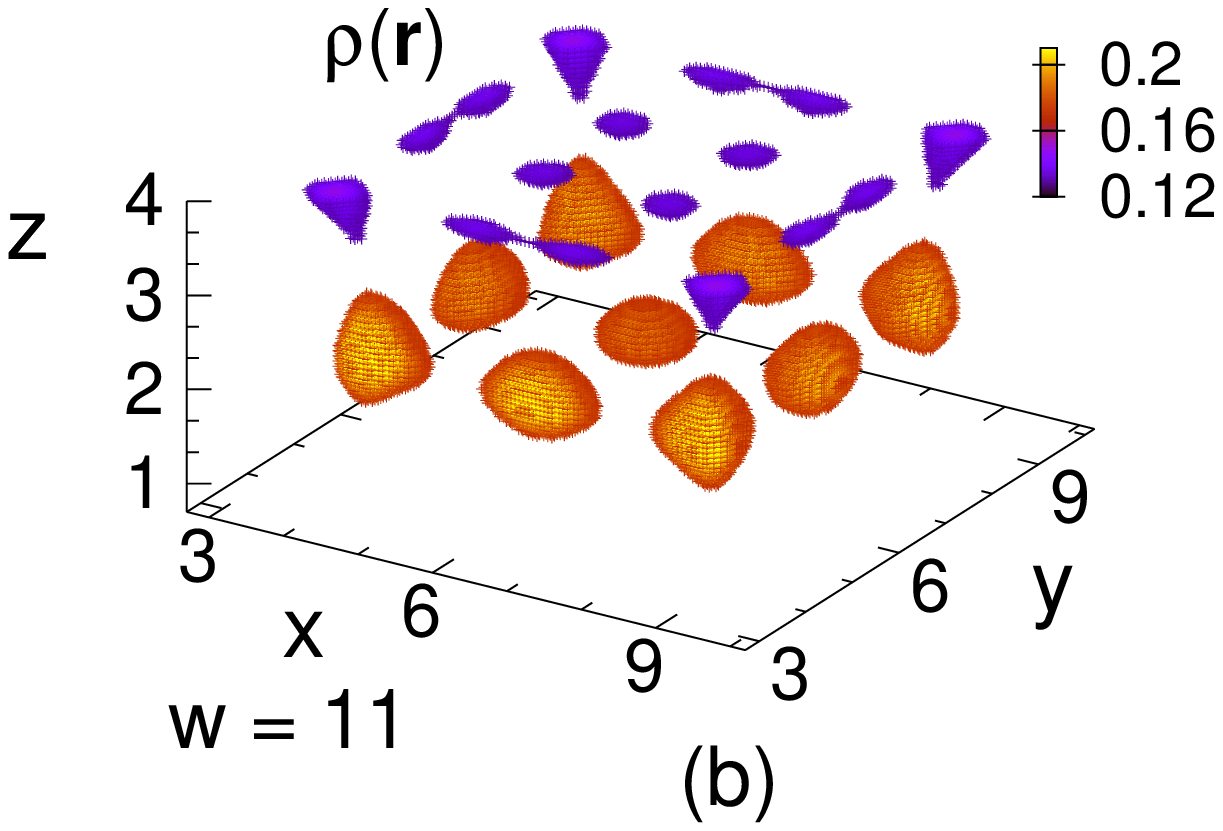}
\hspace{0.6cm}\includegraphics[height=1.6in,width=1.3in]{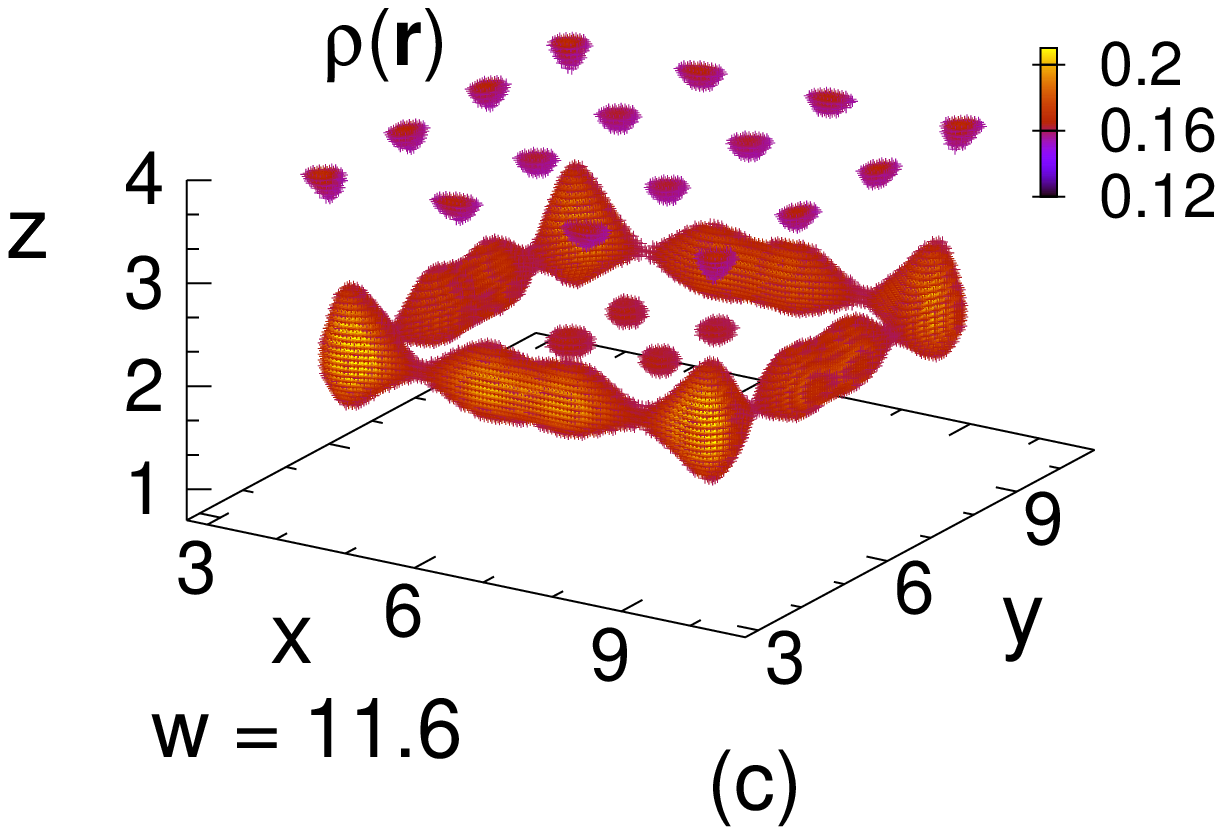}
\hspace{0.6cm}\includegraphics[height=1.6in,width=1.3in]{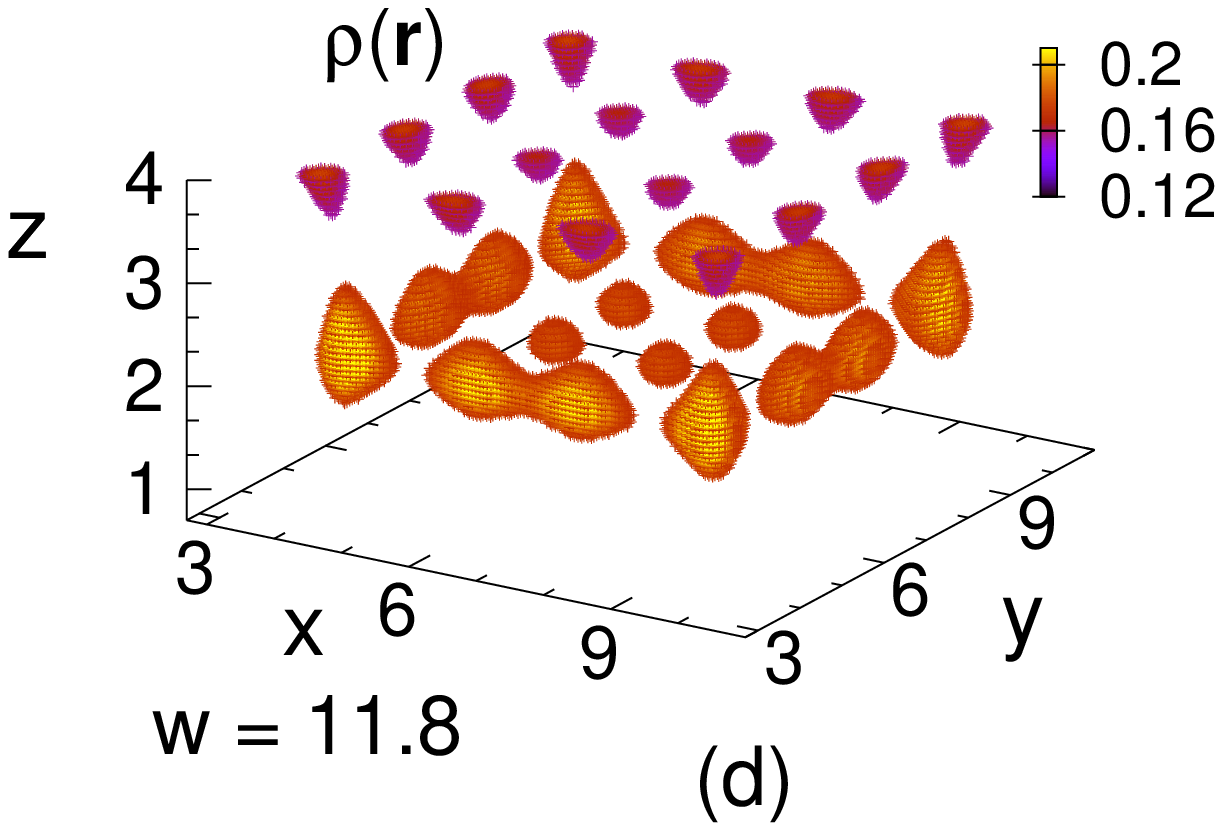}
\vspace{-0.2cm}
\caption{Three-dimensional representation of the fluid structures intermediate between the $3 \times 3$ and
$4 \times 4$ configuration at $\eta=0.42$ and for $D=4.0$. Only the regions with the density higher
than $0.12$ are shown.}
\end{figure*}

In Figs. $5(a)-(f)$, as function of $w$ the evolution of the structure of the fluid in the second 
layer (close to the opening of the pit) is shown. In Figs. $5(g)-(l)$ the emergence of the fluid 
structure in the first layer (deeper in the pit) can be followed upon increasing $w$. 
For the upper layer the formation of the $4 \times 4$ structure occurs already at a smaller width than 
for the lower layer because for the upper layer, due to the rounding of the depletion zone at the 
pit opening,  there is effectively more space available in lateral directions.

Figure $6$ provides a three-dimensional view of the fluid structures formed between the 
fully developed $3 \times 3$ and $4 \times 4$ configurations. As can be inferred from  
Figs. $5$ and $6$, upon varying $w$, the transformation between the two structures proceeds via 
intermediate configurations with broadened density distributions. Presumably this occurs because 
packing effects in the liquid and the constraints imposed by the walls of the pits are 
incommensurate for certain intermediate pit widths. 

\begin{figure*}[ht]
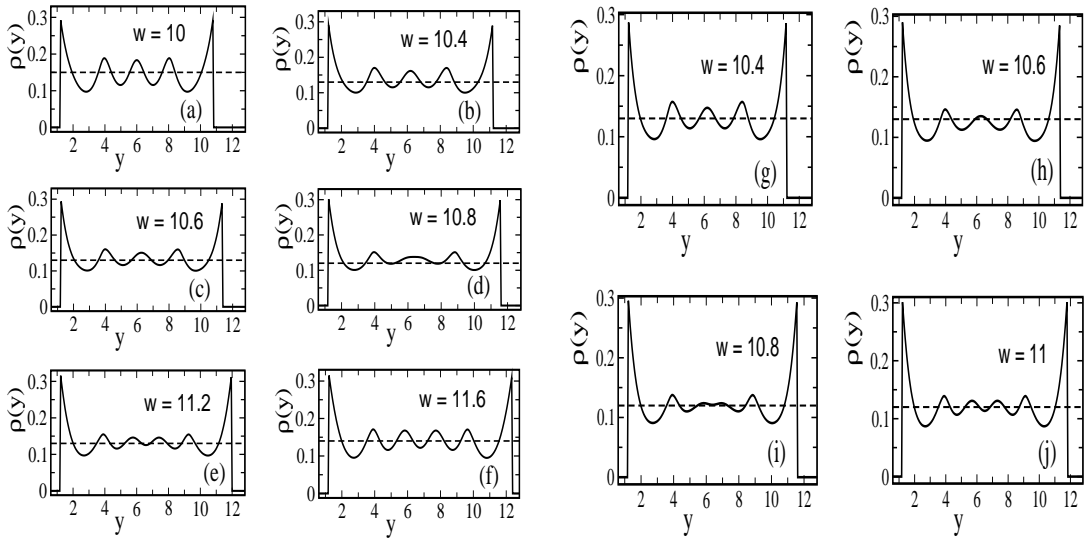


\hspace{-1.4cm}\includegraphics[height=2.8in,width=2.7in]{Fig7_a-f.eps}
\hspace{0.5cm}\includegraphics[height=2.8in,width=2.7in]{Fig7_g-j.eps}

\vspace{-0.3cm}

\caption{Density distributions for $\eta=0.42$ and $D=4.0$ along the line $l_{1}$ ($x=4, z=4$)
((a) - (f)) and the line $l_{2}$ ($x=6, z=4$) ((g) - (j)) in the upper, {\it i.e., } second layer
(see Fig. $4(a)$). The dashed lines indicate the threshold values for $\rho$ leading to Fig. $4$.}
\end{figure*}

\begin{figure*}
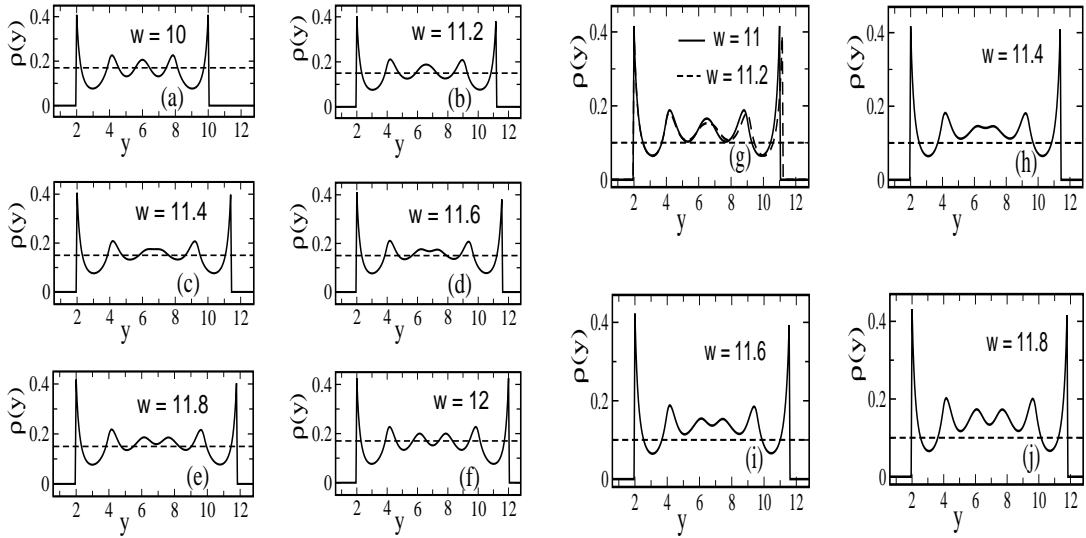


\vspace*{0.50cm}

\hspace{-1.4cm}\includegraphics[height=2.8in,width=2.7in]{Fig8_a-f.eps}
\hspace{0.5cm}\includegraphics[height=2.8in,width=2.7in]{Fig8_g-j.eps}

\vspace{-0.3cm}

\caption{Density distribution for $\eta=0.42$ and $D=4.0$ along the line $l_{3}$ ($x=4, z=2$)
((a) - (f)) and the line $l_{4}$ ($x=6, z=2$) ((g) - (j)) in the lower, {\it i.e., } first layer
(see Fig. $4(b)$).}
\end{figure*}

In order to provide more explicitly quantitative results, we also show density profiles along certain 
selected lines running through the pits. In Fig. $7$ the density profiles in the upper layer along the 
lines $l_{1}$, and $l_{2}$ indicated in Fig. $4$ are shown.  They illustrate how the density profiles vary 
as a function of the width $w$. In Fig. $8$ the density profiles in the lower layer along the lines $l_{3}$ and $l_{4}$ 
indicated in Fig. $4$ are shown. These density profiles reflect the structural changes illustrated in Figs. $3 - 6$
upon varying $w$. 
In order to obtain also a coarse grained description of the liquid inside the pits of volume
$V_{P}$ we consider the average density

\begin{figure*}[ht]

\vspace{1.4cm}

\hspace{0.5cm}\includegraphics[height=2.4in,width=4.8in]{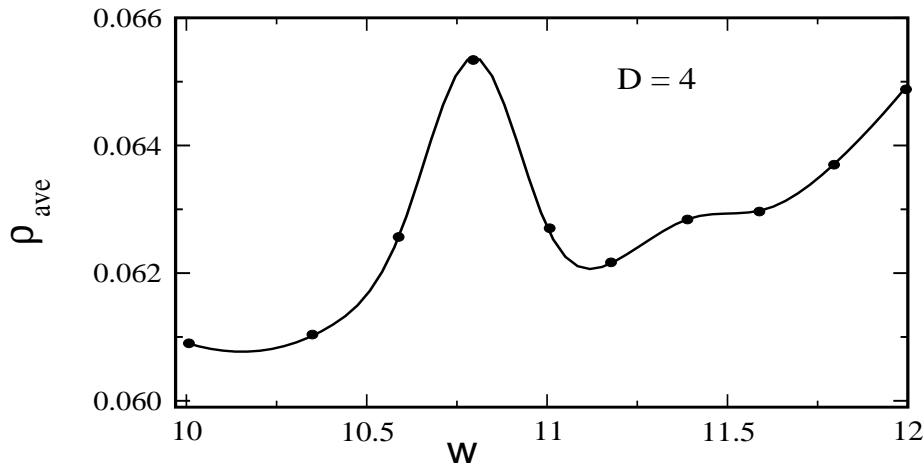}
\vspace{-0.2cm}
\caption{Average density $\rho_{ave}$ (Eq. $(4.2)$) inside the pit as a function of the pit width
$w$ for $D=4$ and $\eta=0.42$ which corresponds to $\rho_{bulk}=0.10$. Accordingly, the average
density inside the pit is below the bulk density, which is expected to be attained for very large $w$.
The solid line smoothly interpolates the discrete data points.}
\end{figure*}


\be
\rho_{ave} = \frac{1}{V_{P}}\int_{V_{P}}d^{3} r \rho{(\bf r)} \approx \frac{1}{N}\sum_{i}\rho(x_{i}, y_{i}, z_{i}),
\ee
where $x_{1}\leq x_{i}\leq x_{2}$, $y_{1}\leq y_{i}\leq y_{2}$, $-R\leq z_{i}\leq D$ (see Fig. $1$)
and $N$ is the total number of grid points inside the pit.
This definition counts also the vanishing density within the depletion zones adjacent 
to the inner walls of the pits (see Figs. $7$ and $8$). The non-monotonous variation of $\rho_{ave}$ 
as a function of the width $w$ of the pits (see Fig. $9$) reflects the transformations of the  liquid 
structure shown in Figs. $5$ and $6$.

For comparison we also show some results obtained for the higher packing fraction $\eta = 0.46$.
In Figs. $10(a)$ and $10(b)$, the liquid densities are presented in vertical cuts for a pit of width 
$w = 10.0$. In Figs. $10(c)$ and $10(d)$ the three-dimensional configurations of the high density 
spots are shown for the different pit widths $w = 9.6$ and $w = 11.6$. The liquid structure observed 
for $\eta=0.46$ is qualitatively similar to the ones obtained for $\eta=0.42$, although at larger values
of $\eta$ the systems attain the $3 \times 3$ or the $4 \times 4$ configuration at smaller pit widths 
(compare Figs. $5$, $6$, $10(c)$, and $10(d)$). The contrast between the densities in the  
high density spots and in the surrounding regions of lower density is more pronounced for the
higher packing fractions. Upon increasing $\eta$ the density inside the high density spots becomes 
much higher whereas the density in the low density regions decreases with increasing packing 
fraction (see Fig. $11$).

\begin{figure*}

\vspace{0.5cm}

\hspace{-0.5cm}\includegraphics[height=1.6in,width=1.3in]{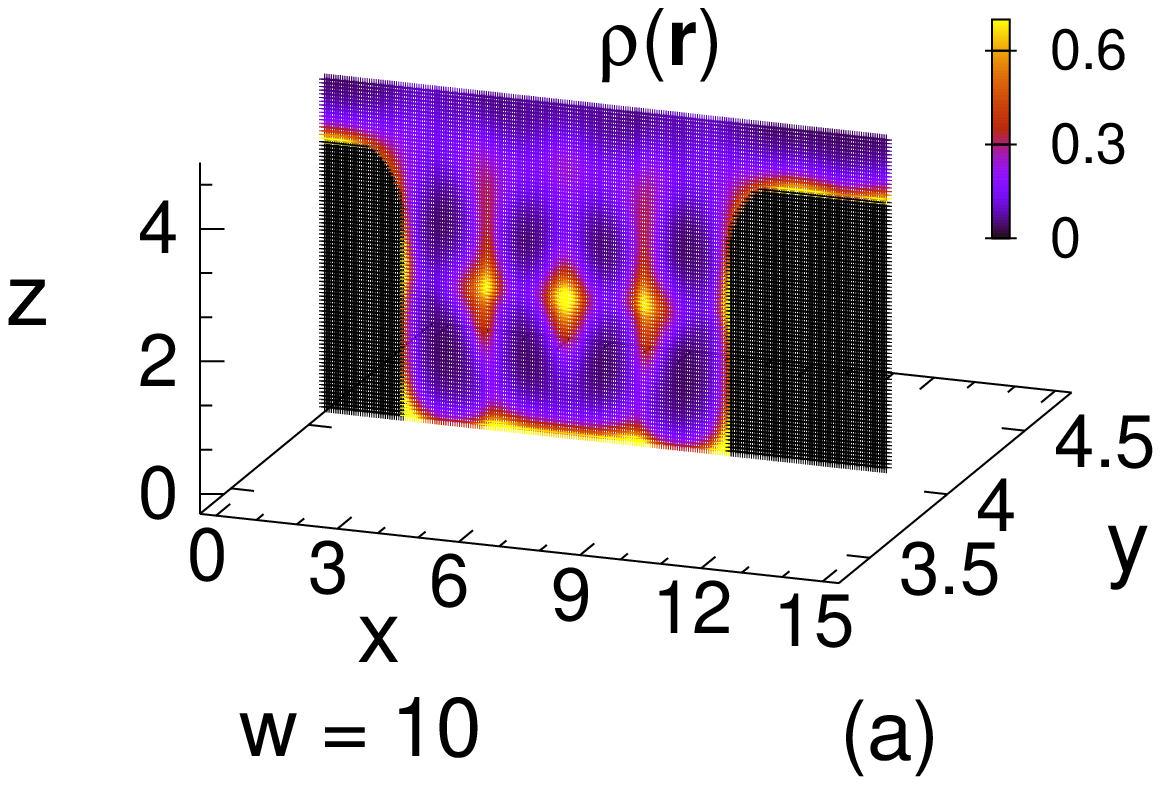}
\hspace{0.60cm}\includegraphics[height=1.6in,width=1.3in]{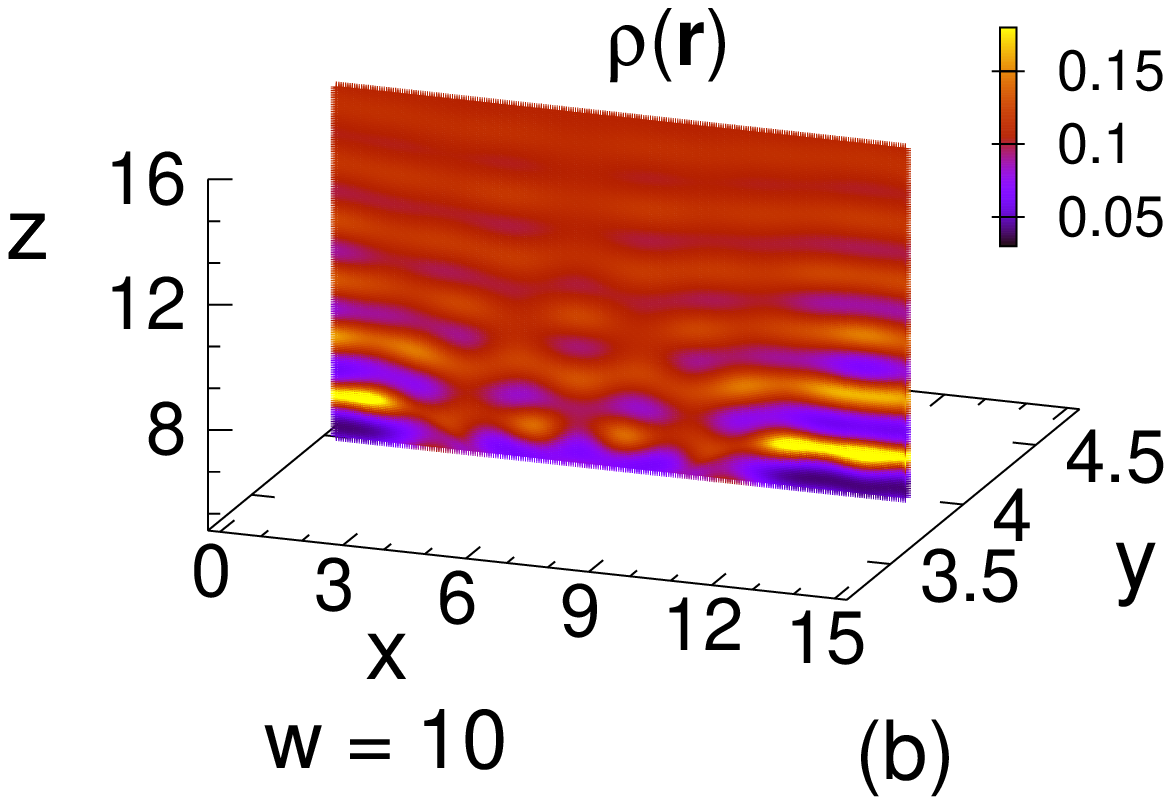}
\hspace{0.60cm}\includegraphics[height=1.7in,width=1.4in]{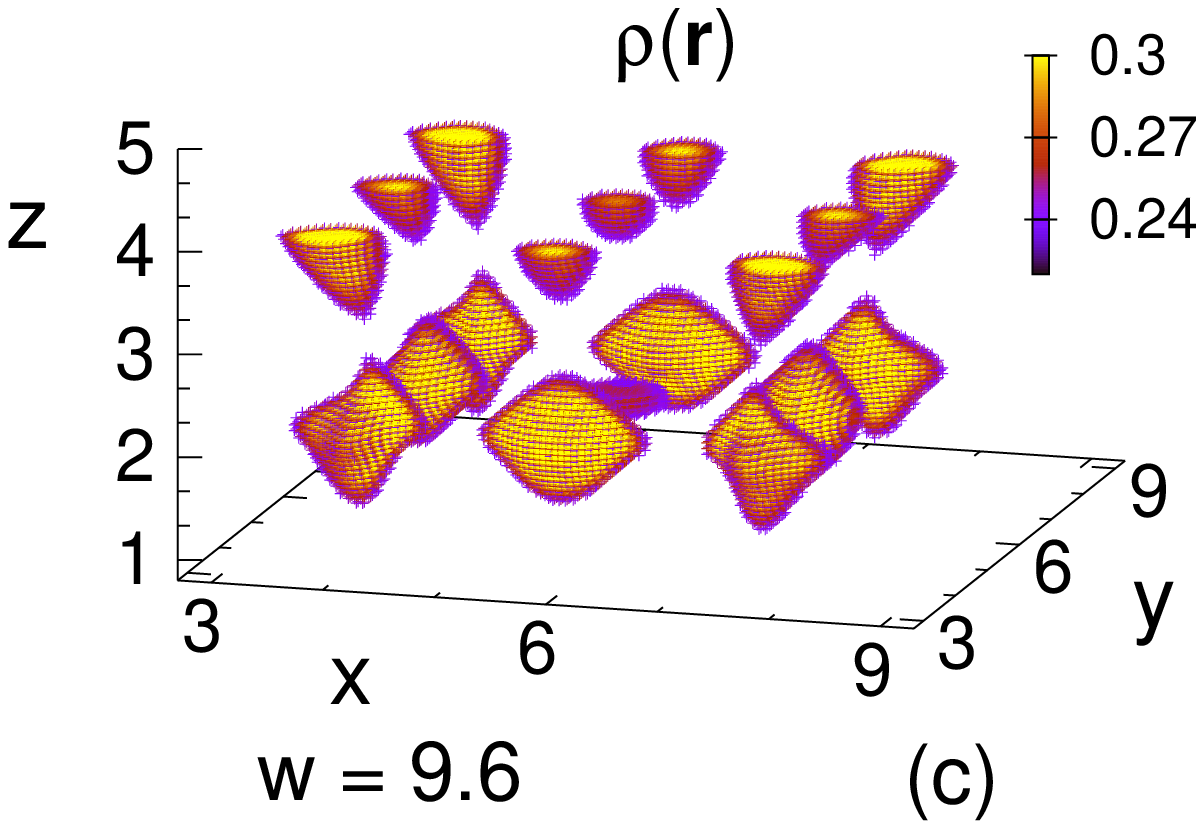}
\hspace{0.60cm}\includegraphics[height=1.7in,width=1.4in]{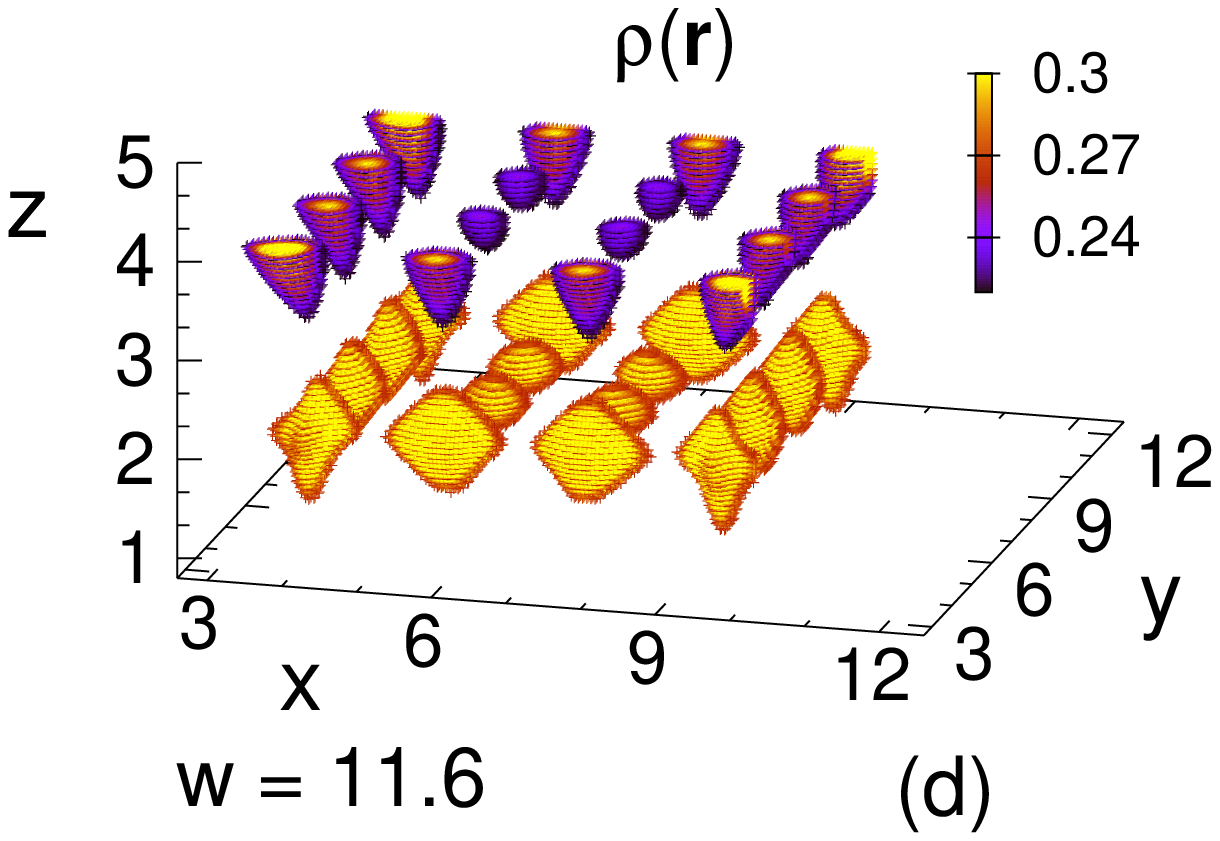}

\caption{Vertical cut ($xz$ plane at $y=4$) through the pit ((a) and (b)) and the three-dimensional
representation ((c) and (d)) of the fluid structure inside the pit for $\eta=0.46$, $D=4.0$
, and for two commensurate widths $w=9.6$ and $w=11.6$. The vertical cut goes through the
position of the density maximum adjacent to the density peak at the front side of the pit.
In the three-dimensional representation for the bottom layer only the regions with the density
higher than $0.25$ are shown whereas the threshold density for the upper layer is chosen to be $0.22$.
The high-density regions next to the walls are not shown.}
\end{figure*}

\begin{figure*}[ht]

\vspace{1.0cm}

\hspace{-1.2cm}\includegraphics[height=2.5in,width=4.8in]{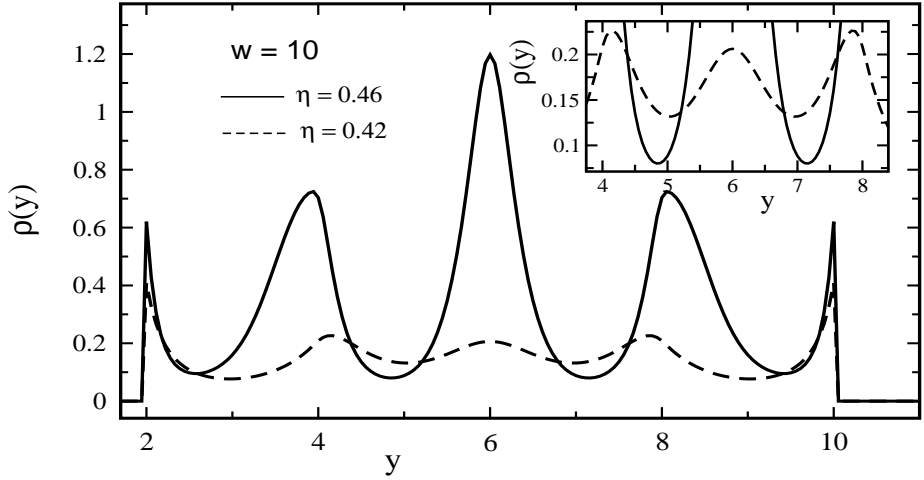}
\vspace{-0.2cm}
\caption{Same as Fig. $8(a)$ for the packing fraction $\eta=0.46$, compared with the case
$\eta=0.42$.}
\end{figure*}

\subsection{Structure of the liquid as a function of the depth of the pits}
In this subsection we study how the liquid structure within the pits evolves as a function of 
their depths $D$ for a constant width $w=10.0$. To this end the depth $D$ has been varied between 
$D=4.0$ and $D=11.0$. The liquid structure inside the pits for $D=4.0$ and $w=10.0$ has been 
described already in Subsec. III A. It is characterized by two layers stacked upon each other
in the normal $z-$ direction, each layer forming a $3 \times 3$ array of high density spots. 
Between the spots the liquid density is considerably lower than the density inside the spots.
The spots in the upper ({\it i.e., } second layer) are exactly stacked upon those of the bottom 
({\it i.e., } first layer),  separated by regions of low density liquid. Due to this aligned stacking 
one could also state that the regions of high density form a columnar structure, although for $D=4.0$
this column consists of only two separate spots.
Upon increasing the depth of the pit, between these spots connecting necks are formed and eventually 
additional spots bulge out in vertical direction. Thus the high density regions form 
columns with narrow constrictions at certain depths of the pit. Depending on the depth these
columns may disintegrate partially or completely into separated spots stacked upon each other
This structure of the columns reflects the layering in the vertical direction. This structural
transformation as a function of the depth $D$ is illustrated in Fig. $12$ for the
central column (Figs. $12(a)$-$(d)$) within the $3 \times 3$ configuration and for a
corner column (Figs. $12(e)$-$(h)$). At $D=7.4$, the central column consists of three separated or almost 
separated spots. As the depth is increased the central column eventually develops $4$ separated 
spots (at about $D=9$). The corner columns show a similar evolution of new spots for those parts 
deeper inside the pits whereas in these columns the density distribution towards 
the pit opening is rather smeared out (see Figs. $12(f)$ and $(h)$). The columns facing the 
walls (not shown) show a behavior which is intermediate between that of the corner
columns and of the central column.

\begin{figure*}[ht]

\hspace{-2.65cm}\includegraphics[height=1.4in,width=1.0in]{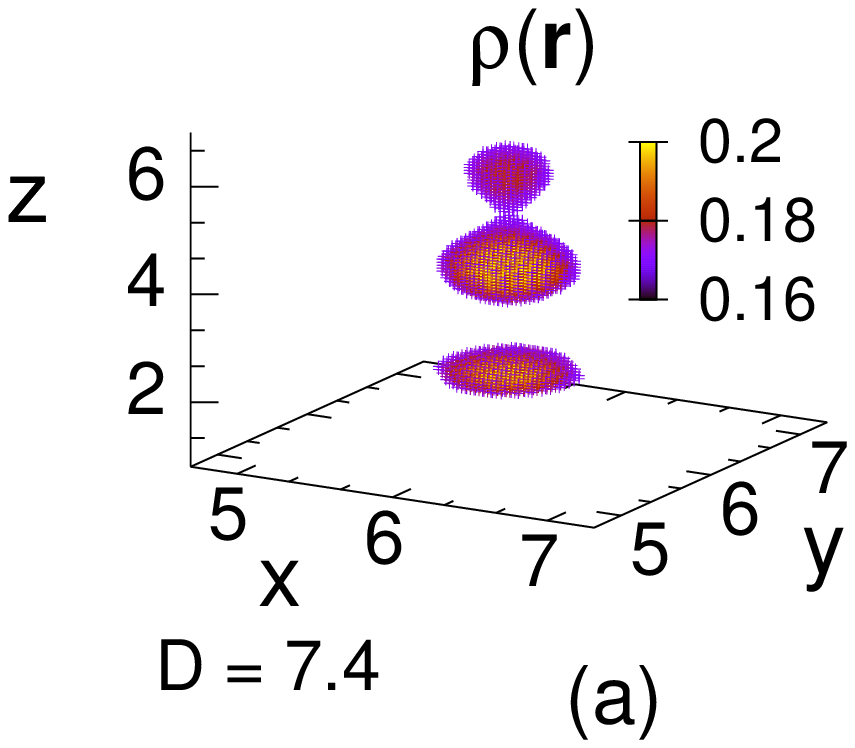}
\hspace{0.70cm}\includegraphics[height=1.4in,width=1.0in]{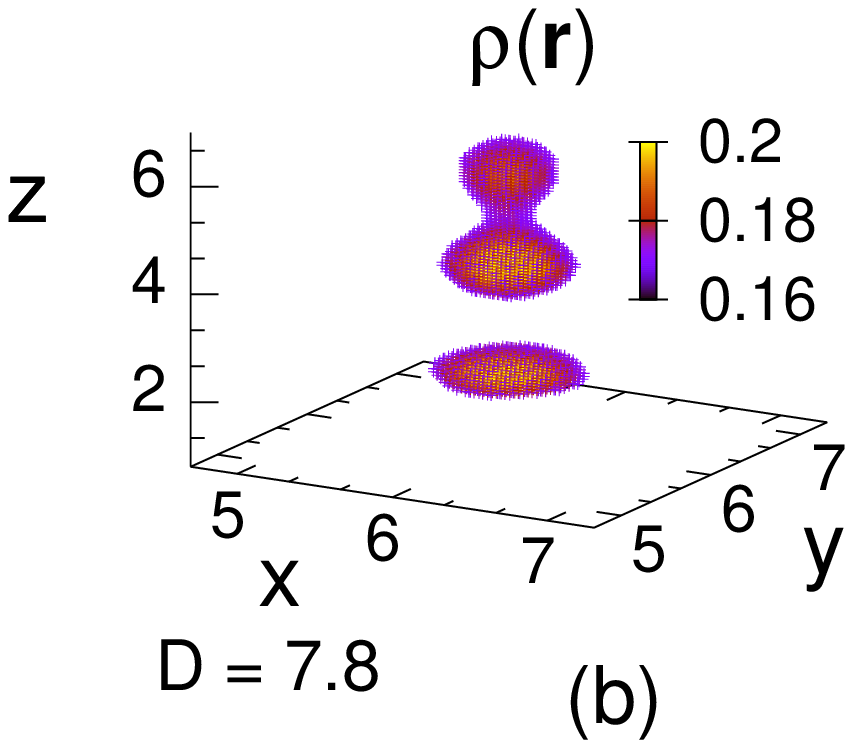}
\hspace{1.70cm}\includegraphics[height=1.4in,width=1.0in]{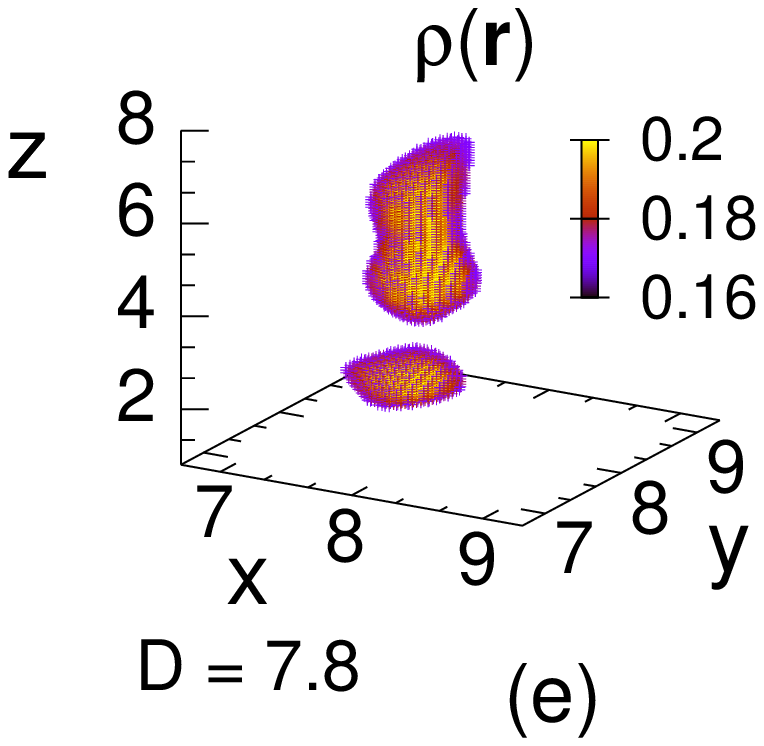}
\hspace{0.70cm}\includegraphics[height=1.4in,width=1.0in]{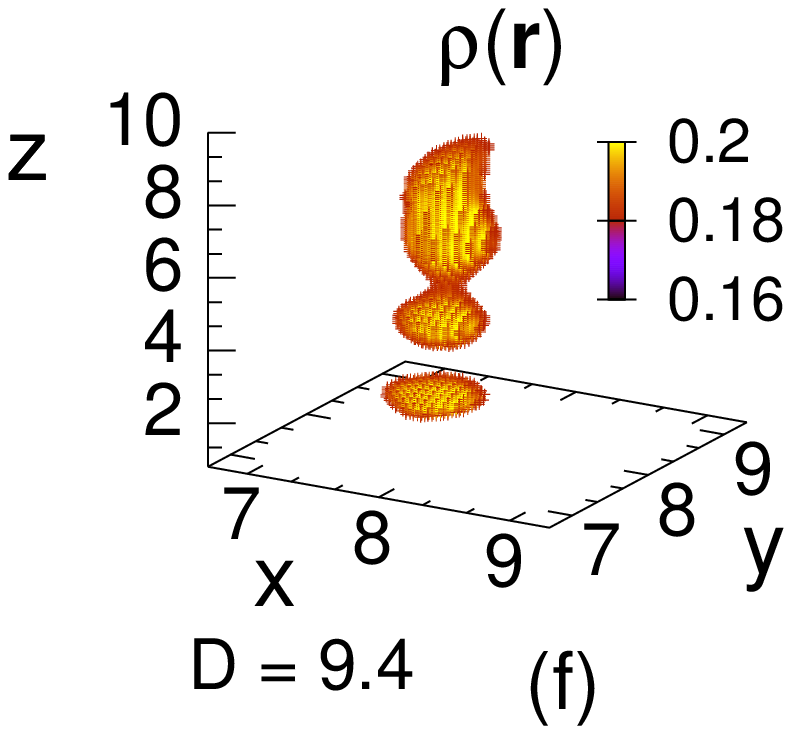}

\vspace{0.5cm}

\hspace{-2.65cm}\includegraphics[height=1.5in,width=1.0in]{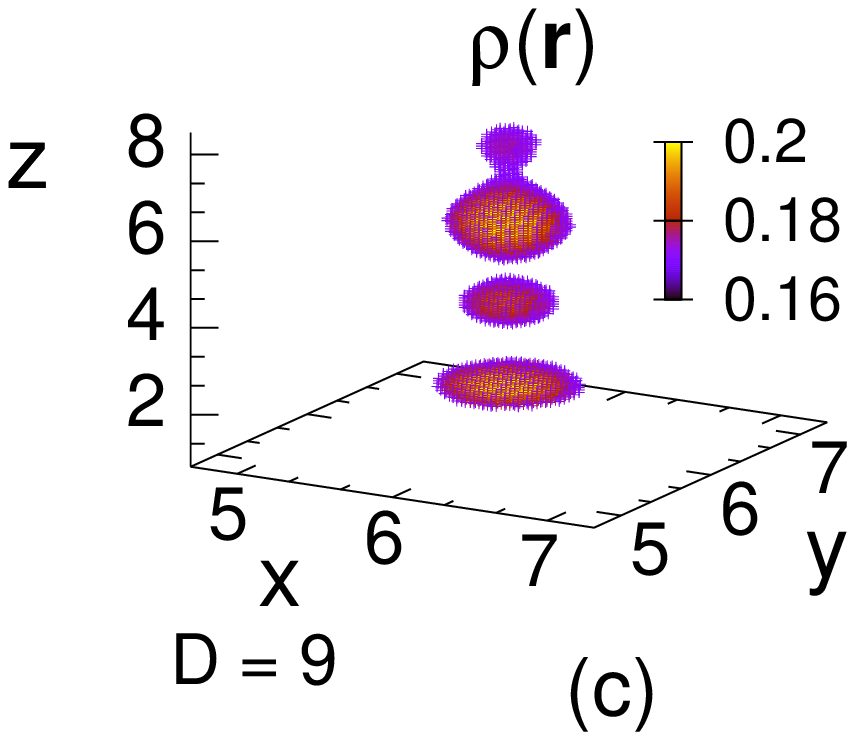}
\hspace{0.70cm}\includegraphics[height=1.5in,width=1.0in]{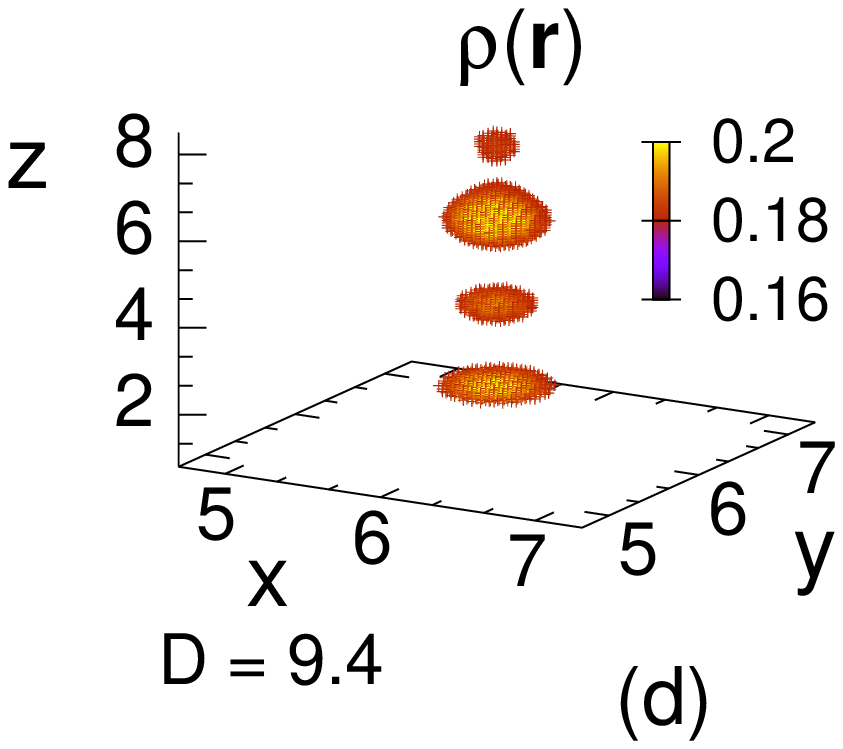}
\hspace{1.70cm}\includegraphics[height=1.5in,width=1.0in]{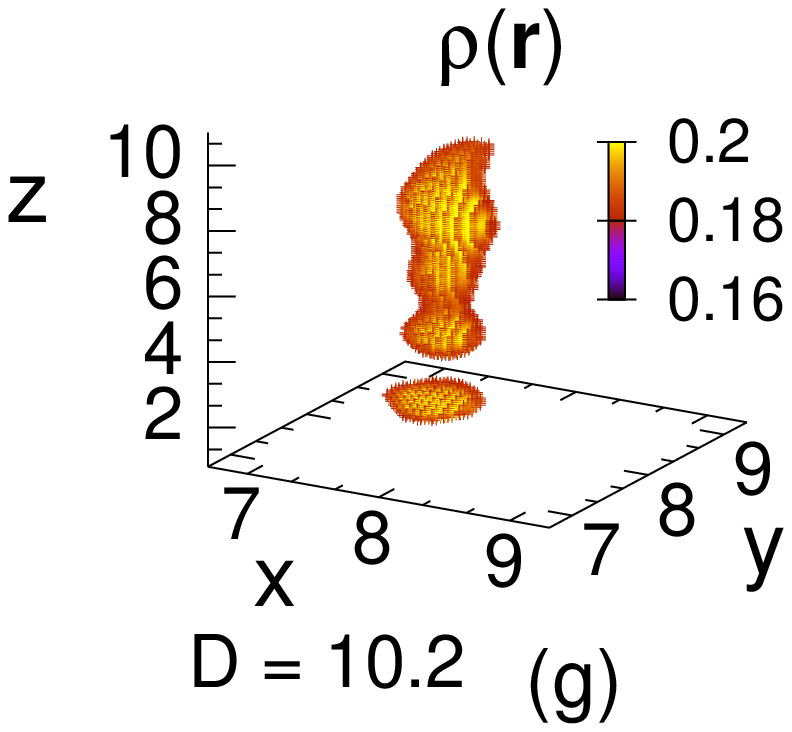}
\hspace{0.70cm}\includegraphics[height=1.5in,width=1.0in]{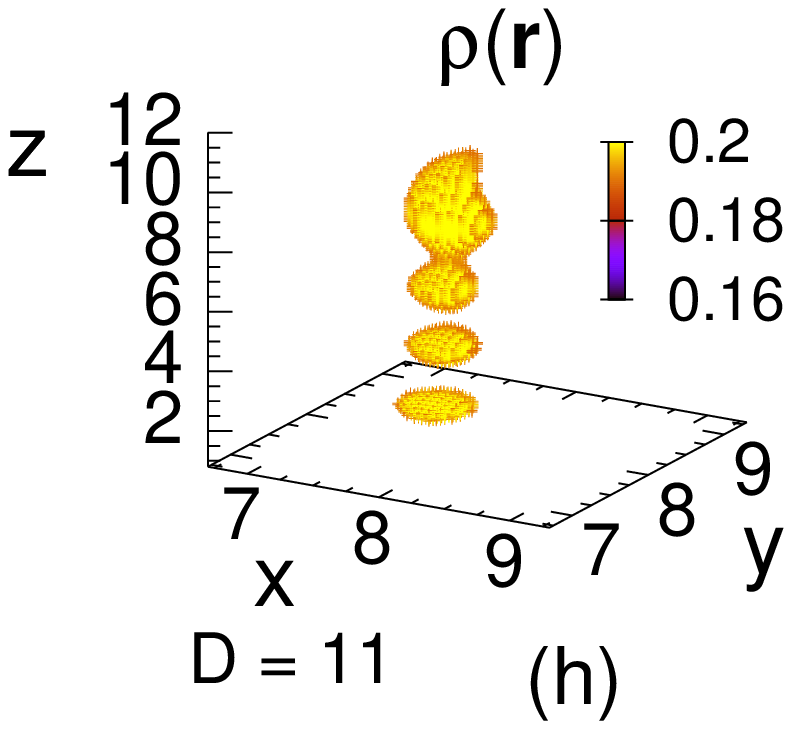}
\vspace{-0.2cm}
\caption{Transformations of the  structure of the liquid inside the pit upon increasing the  depth
$D$ for constant width $w=10.0$ at $\eta=0.42$. The vertical position of the floor of the pit is fixed
at $z_{1}=-1.0$ and $z_{2}$ is varied in order to create pits of various depths.
Panels $(a)-(d)$ show the center column (see Figs. $3(a)$) and panels $(e) - (h)$
show a corner column. Upon increasing $D$, additional spots emerge via a budding-like process.}
\end{figure*}

\begin{figure*}
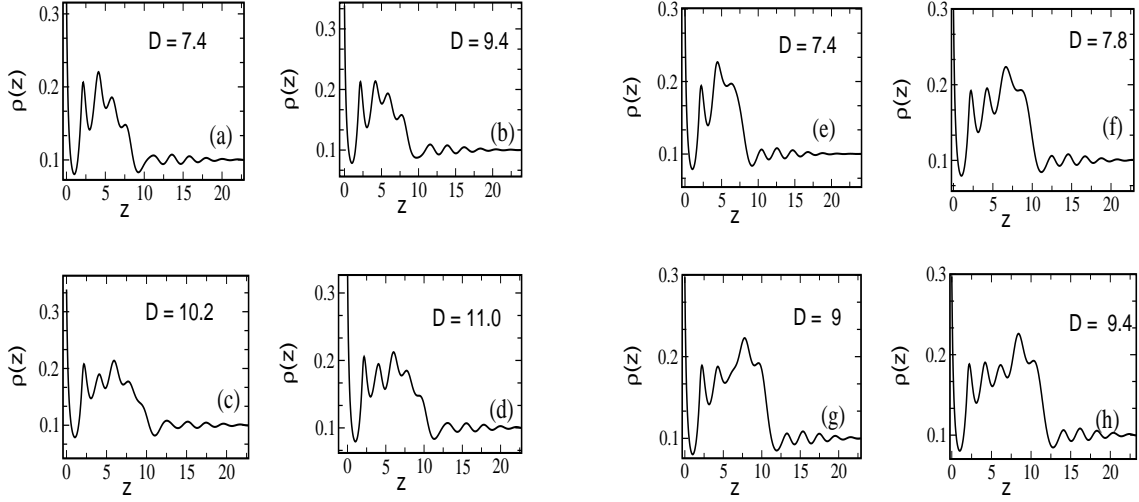

\hspace{-2.0cm}\includegraphics[height=2.6in,width=2.7in]{Fig13_a-d.eps}
\hspace{1.2cm}\includegraphics[height=2.6in,width=2.7in]{Fig13_e-h.eps}
\vspace{-0.2cm}
\caption{ Density profiles $\rho(z)$ for $\eta=0.42$ and $w=10.0$ along the $z-$axis
at ($x=6, y=6$) for the central column ((a)-(d)) and at ($x=4, y=4$)
for a corner column ((e)-(h)); compare Fig. $12$.}
\end{figure*}

\begin{figure*}[ht] 
\vspace{0.2cm}
\includegraphics[height=3.0in,width=4.8in]{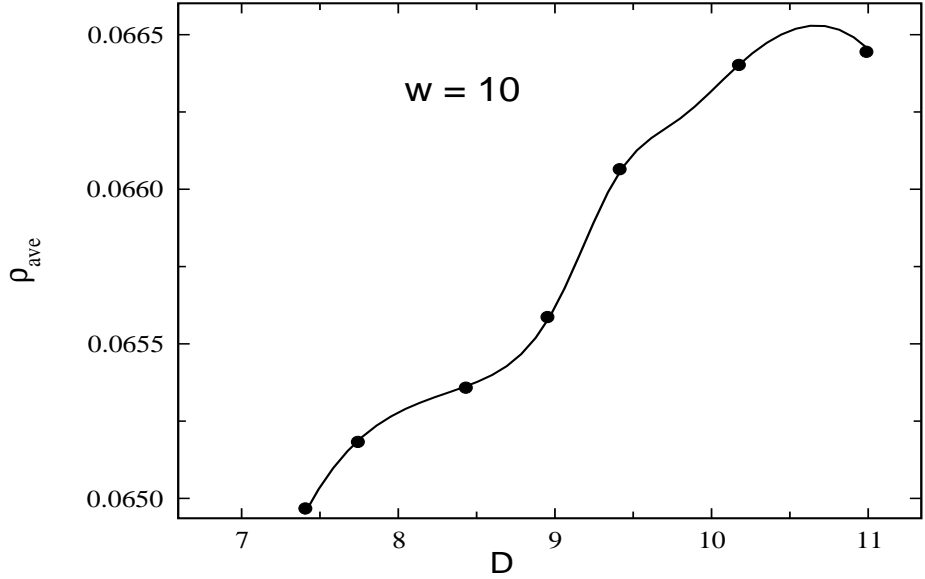}
\vspace{-0.2cm}
\caption{ Average density $\rho_{ave}$ (Eq. $(4.2)$) inside the pit as a function of the depth
$D$ for $w=10.0$ and $\eta=0.42$ which corresponds to $\rho_{bulk}=0.10$. The solid line smoothly interpolates
the discrete data points.}
\end{figure*}


For a better quantitative description we also show density profiles along the axis of the 
central column (see Figs. $13(a)-(d)$) and along the axis of a corner column 
(see Figs. $13(e)-(h)$) for the $3 \times 3$ configuration.   
The small density oscillations at large $z$ values are due to layering above the pits. 
In Fig. $14$, we show the average density inside the pits as a function of their depths.
We find a monotonous increase of $\rho_{ave}$ upon increasing the pit depth.
\section {\bf Summary}
We have studied the structure of a Lennard-Jones type liquid in contact with a sculptured solid wall  
interacting with the fluid particles via a hard repulsion and a van der Waals attraction.
The wall is endowed with nanopits of square cross section. The walls
forming that structure are taken to be piecewise flat. Inside the nanopits three-dimensional 
localization of the liquid can be observed with high density spots separated from each other 
in all spatial directions by regions with considerably lower number density. The onset of this 
localization occurs already at packing fractions much lower than those of the liquid at bulk 
liquid-solid coexistence. For suitably chosen widths and depths of the pits such that they are
commensurate with packing requirements, the high density spots are spatially compact and form 
simple cubic lattices or they are at least ordered on square lattices
in certain planes. The number of spots depends on the width and the depth of the pit. Too strong 
deviations from the commensurate widths lead to a broadening of the density distribution and 
to the formation of  bridges between high density spots. We have followed the evolution of these 
structures as the width of the pit is varied from one commensurate value to the next one. 
Above the pits mainly layering is observed in the vertical direction. The layers are 
distorted above the pit opening, but these distortions die out within a few molecular diameters.  

The degree of localization inside the pit becomes more pronounced as the packing fraction increases. 
Qualitatively, however, the structures remain to be similar. For packing fractions, which 
correspond to the liquid phase in the bulk, within the pits no region with genuine crystalline order 
has been observed for the fluid-wall interactions, pit geometries, and sizes 
studied here.

Our results also hold for colloidal suspensions with pits comparable in size with
that of the colloids. In that case
the pit dimensions and the length scale characterizing the structures discussed in the
present work are not in the $nm$ range but may be scaled-up to much larger length 
scales. Of course the effective interactions among the colloidal particles and
between the colloidal particles and the wall have to be tuned such that agglomeration or complete sedimentation are 
prevented. In particular for molecular liquids, these structures at the sculptured surface-liquid 
interface are in situ experimentally accessible by small-angle X-ray scattering \cite{Checco}.

\newpage

\begingroup

\end{document}